# Propagation of weakly stretched premixed spherical spray flames in localized homogeneous and heterogeneous reactants

Qiang Li, Huangwei Zhang[*] and Chang Shu

*Department of Mechanical Engineering, National University of Singapore, 9 Engineering Drive 1, Singapore 117576, Republic of Singapore*

**Abstract**

Propagation of weakly stretched spherical flames in partially pre-vaporized fuel sprays is theoretically investigated in this work. A general theory is developed to describe flame propagation speed, flame temperature, droplet evaporation onset and completion locations. The influences of liquid fuel and gas mixture properties on spherical spray flame propagation are studied. The results indicate that the spray flame propagation speed is enhanced with increased droplet mass loading and/or evaporation heat exchange coefficient (or evaporation rate). Opposite trends are found when the latent heat is high, due to strong evaporation heat absorption. Fuel vapor and temperature gradients are observed in the post-flame evaporation zone of heterogeneous flames. Evaporation completion front location considerably changes with flame radius, but the evaporation onset location varies little relative to the flame front when the flame propagates. For larger droplet loading and smaller evaporation rate, the fuel droplet tends to complete evaporation behind the flame front. Flame bifurcation occurs with high droplet mass loading under large latent heat, leading to multiplicity of flame propagation speed, droplet evaporation onset and completion fronts. The flame enhancement or weakening effects by the fuel droplet sprays are revealed by enhanced or suppressed heat and mass diffusion process in the pre-flame zone. Besides, for heterogeneous flames, heat and mass diffusion in the post-flame zone also exists. The mass diffusion for both homogeneous and heterogeneous flames is enhanced with decreased Lewis number. The magnitude of Markstein length is considerably reduced with increased droplet loading. Moreover, post-flame droplet burning behind heterogeneous flame influences the flame propagation speed and Markstein length when the liquid fuel loading is relatively low.

---

[*] Corresponding author. Email: huangwei.zhang@nus.edu.sg, Tel: +65 6516 2557.

# 1  Introduction

Liquid fuel is predominantly used in many combustion applications, e.g. aero-engines and rocket engines. High-efficiency and low-emission spray combustion devices are in high demand nowadays. To achieve this, it is significant to first clarify the fundamental aspects of two-phase combustion, and one of them is flame initiation and propagation in sprayed fuel droplets.

The effects of fuel droplet properties on spray flame propagation have been extensively studied. For instance, Mizutani et al. [1,2] observe that adding kerosene droplets intensidies the propane/air combustion. However, it is also found that there exist appropriate droplet quantities for combustion enhancement [1]. Moreover, Myers and Lefebvre [3] study six liquid fuel sprays and find that the flame speed is inversely proportional to mean droplet diameter above some critical value, and increases with overall equivalence ratio. Hayashi and Kumagai [4] observe that the flame speed decreases with liquid fuel loading in overall fuel-lean mixtures. Nomura et al. [5–7] find that the flame speed of fuel (methanol and ethanol) droplet−vapor−air mixtures exceeds that of premixed gaseous of the same total equivalence ratio in the fuel-lean and fuel-rich regions of the total equivalence ratio. Moreover, Atzler et al. [8] observe that the burning rates of *iso*-octane/air aerosols are strongly affected by droplet diameter when the equivalence ratio is high. Similar results are also achieved by Bradley et al [9], through analyzing equivalence ratio and droplet size effects on flame propagation speeds of *iso*-octane and ethanol aerosols. It is also seen by Neophytou and Mastorakos [10] that the high flame speed of *n*-heptane and *n*-decane is achieved with small droplet diameters and long residence time under fuel-lean condition.

The foregoing influences of the liquid fuel sprays on flame propagation are related to the droplet behaviors, e.g. movement, heating and evaporation [11,12]. It is observed through spray combustion experiments [4,8,13–17] that small-sized droplets can complete evaporation in the preheating zone or immediately around the flame front, but relatively large droplets penetrate though the flame front and continue vaporizing in the post-flame zone. Three spray flame propagation modes are identified from the droplet-level OH* chemiluminescence and OH-PLIF images by de Oliveira and Mastorakos [13], including droplet propagation, inter-droplet propagation and gaseous-like propagation modes. Furthermore, Thimothée et al. [17] study the passage of liquid droplets through a spherical flame and find that droplet size and inter-droplet distance are the most important controlling factor. These peculiar phenomena have also been observed in DNS of spray combustion [18,19].

However, how the droplet distribution and flame propagation intrinsically interact with each other is still not well understood. This is partly caused by the comprehensive (hence complicated) interphase coupling in spray flames, and therefore it is challenging to underpin the mechanism behind the observed flame / droplet behaviors. Theoretical analysis is deemed a powerful tool for two-phase combustion studies, since it can retain or isolate the most relevant factors in the studied problem, thereby highlighting physical relevance and ensuring conclusion generality. For instance, Lin et al. [20,21] develop a theoretical model for fully and partially pre-vaporized burning sprays. Greenberg [22,23] derives an evolution equation for a laminar flame propagation into fuel spray cloud, considering finite-rate evaporation and droplet drag effects. However, in the above studies [20–23], the flame−droplet interactions are not studied. Recently, Han and Chen [24] further examine the influences of finite-rate evaporation on spray flame propagation and ignition, and find that flame propagation speed, Markstein length and minimum ignition energy are strongly affected by droplet loading and evaporation rate. Nonetheless, in their work, the droplets are assumed to be distributed in the full burned and unburned areas. With a more general theory, Li et al. [25] consider the droplet evaporation completion before or after a steadily propagating planar spray flame front and find that differentiated droplet distributions have significant importance on flame propagation. Zhuang and Zhang [26] theoretically analyze consistently varying droplet distributions in a propagating spherical flame, but only fine water sprays are studied. Therefore, the interactions between liquid fuel sprays and propagating spherical flames merit further investigation.

In this work, we aim to conduct theoretical analysis on propagation of spherical spray flame in partially pre-vaporized fuel sprays. Dynamic droplet distributions with a propagating flame front will be described in our model, characterized by evolving droplet evaporation onset and completion locations. This leads to localized homogeneous and heterogeneous reactants at the flame front. The influences of liquid fuel and gas mixture properties will be examined, including evaporation heat transfer coefficient (or evaporation rate), droplet mass loading, latent heat of vaporization and Lewis number. The rest of the paper is structured as below. The physical and mathematical model are presented in Section 2. The analytical solutions are listed in Section 3. The results will be discussed in Section 4. Section 5 closes the paper with main conclusions.

## 2  Physical and mathematical model

### 2.1  Physical model

In this work, one-dimensional spherical flame in partially pre-vaporized fuel sprays will be studied. Two general scenarios are considered, with evaporating droplets: (1) in both pre- and post-flame zones, (2) in pre-flame zone only. The sketches of their physical models are shown in Figs. 1(a) and 1(b), respectively. There are three different characteristic locations for liquid and gas phases in our model, including gaseous flame front ($R_f$), droplet evaporation onset ($R_v$) and completion ($R_c$) fronts. Specifically, $R_v$ corresponds to the location where the droplet starts to evaporate. The droplets are just heated up to boiling temperature, and behind this front (i.e. $R < R_v$), the droplet temperature maintains the boiling temperature and evaporation continues [24,26–29]. The evaporation onset front $R_v$ is always before the flame front $R_f$, indicating that onset of droplet vaporization spatially precedes the gaseous combustion. Moreover, $R_c$ denotes the location at which all the droplets are critically vaporized. When $R < R_c$, no droplets are left and hence their effects on the gaseous flame diminish.

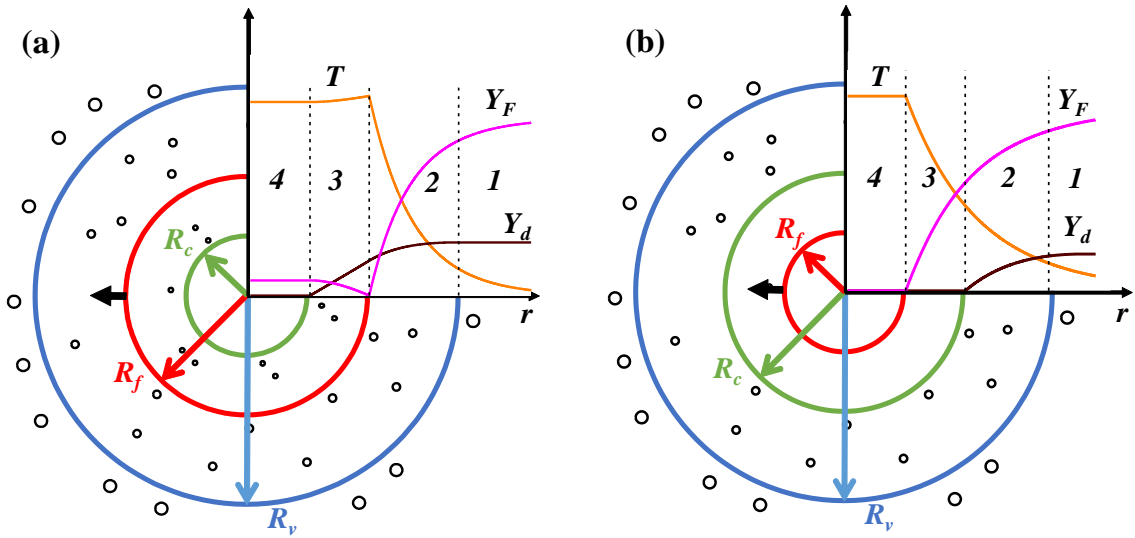

Fig.1 Schematic of outwardly propagating spherical flame in liquid fuel mists: (a) heterogeneous flame, (b) homogeneous flame. Circle: fuel droplet. Red line: flame front ($R_f$); green line: evaporation completion front ($R_c$); blue line: evaporation onset front ($R_v$). Black arrow: flame propagation direction.

In Fig. 1(a), the evaporation completion front lies after the flame front (i.e. $R_c < R_f < R_v$). The local mixture around the flame front $R_f$ is composed of gaseous vapor and evaporating fuel droplets. In Fig. 1(b), the evaporation completion front is located before the flame front (i.e. $R_f < R_c < R_v$), and the mixture around the flame front is purely gaseous, since all the droplets have been gasified into vapor there. For brevity, hereafter, we term the first and second cases as heterogeneous (abbreviated as "HT") and homogeneous ("HM") flames, respectively. As shown in Fig. 1, for both flames, zone 1 represents the pre-vaporization zone before $R_v$. Zone 2 indicates pre-flame evaporation zone before $R_f$ for heterogeneous flame and before $R_c$ for homogeneous flame. As for zone 3, it represents post-flame evaporation zone before $R_c$ for heterogeneous flame, and pre-flame zone without evaporation for homogeneous flame. Meanwhile, zone 4 is the post-flame zone without evaporation for both flames. In the following, we will develop a general theory to describe propagation and transition of homogeneous and heterogeneous spherical spray flames, considering consistently evolving fuel droplets with the moving reaction front.

## 2.2 Governing equation

For gaseous flames, the well-known diffusive-thermal model [30,31] is adopted, with which the thermal and transport properties (e.g. density, thermal conductivity and heat capacity) are assumed to be constant and the convection flux is absent. This model has been used in numerous studies on gaseous and two-phase flames [24,26,28,29,32,33]. One-step chemistry is considered, i.e. $F + O \rightarrow P$, with $F$, $O$ and $P$ being fuel vapour, oxidizer and product, respectively. Globally fuel-lean mixture (i.e. total equivalence ratios of fuel vapor and droplets are below unity) is studied in this work. Therefore, the equations for gas temperature and fuel mass fraction are

$$\tilde{\rho}_g \tilde{C}_{p,g} \frac{\partial \tilde{T}}{\partial \tilde{t}} = \frac{1}{\tilde{r}^2} \frac{\partial}{\partial \tilde{r}} \left( \tilde{r}^2 \tilde{\lambda}_g \frac{\partial \tilde{T}}{\partial \tilde{r}} \right) + \tilde{q}_c \widetilde{\omega}_c - \tilde{q}_v \widetilde{\omega}_v + \alpha \tilde{q}_c \widetilde{\overline{\omega}}, \qquad (1)$$

$$\tilde{\rho}_g \frac{\partial \tilde{Y}_F}{\partial \tilde{t}} = \frac{1}{\tilde{r}^2} \frac{\partial}{\partial \tilde{r}} \left( \tilde{r}^2 \tilde{\rho}_g \tilde{D}_F \frac{\partial \tilde{Y}_F}{\partial \tilde{r}} \right) - \widetilde{\omega}_c + \widetilde{\omega}_v - \alpha \widetilde{\overline{\omega}}, \qquad (2)$$

where the tilde symbol ~ is used to indicate that the variables are dimensional. $\tilde{t}$ and $\tilde{r}$ are respectively the temporal and spatial coordinates. $\tilde{T}$, $\tilde{\rho}_g$, $\tilde{C}_{p,g}$, and $\tilde{\lambda}_g$ are the gas temperature, density, heat capacity and thermal conductivity, respectively. $\tilde{Y}_F$ and $\tilde{D}_F$ are the mass fraction and molecular diffusivity of the fuel. $\tilde{q}_c$ is the reaction heat release per unit mass of the fuel. $\tilde{q}_v$ is the latent heat of vaporization of liquid fuel, whilst $\tilde{\omega}_v$ is the evaporation rate of the fuel droplet. The chemical reaction rate $\tilde{\omega}_c$ in Eq. (1) takes the Arrhenius form

$$\tilde{\omega}_c = \tilde{\rho}_g \tilde{A} \tilde{Y}_F \, exp(-\tilde{E}/\tilde{R}^0 \tilde{T}). \qquad (3)$$

Here, $\tilde{A}$ is the pre-exponential factor, $\tilde{E}$ is the activation energy for the reaction, and $\tilde{R}^0$ is the universal gas constant.

In Eqs. (1) and (2), $\alpha$ is an indicator for post-flame droplet burning in heterogeneous flames [24]. Specifically, $\alpha = 0$ indicates that droplet burning in the post-flame zone is neglected, practically corresponding to the situations, e.g. when the local equivalence ratio is beyond the flammability limit and/or the vapor/oxidizer is not well mixed [13]. On the contrary, when $\alpha = 1$, we assume that the fuel vapor is totally consumed by the local diffusion combustion surrounding individual droplets in the post-flame zone [13]. In this case, the last terms in Eqs. (1) and (2) are sink terms, only applicable for the post-flame evaporation zone (zone 3 in Fig. 1a) of the heterogeneous flame. The droplet burning term $\tilde{\bar{\omega}}$ reads

$$\tilde{\bar{\omega}} = \tilde{\omega}_v H(\tilde{R}_f - \tilde{r}). \qquad (4)$$

$H(\cdot)$ is the Heaviside function. Note that when $\alpha = 1$, $\tilde{\bar{\omega}} = \alpha \tilde{\omega}_v$ is valid at the right side of Eq. (2), which implies that all the vaporized fuel behind the flame front $\tilde{R}_f$ is reacted with the heat release of $\tilde{q}_c \tilde{\bar{\omega}}$ in Eq. (1).

For liquid fuel droplets, we assume that they are uniformly monodispersed in the initial pre-vaporized fuel/air mixture. The droplets are spherical and their properties (e.g. density and heat capacity) are assumed to be constant. Due to the dilute droplet concentration, the inter-droplet collisions are not considered and therefore the diffusion of liquid droplets can be neglected. The above assumptions are also used in previous theoretical work on two-phase flames [20,34–36]. Furthermore, in zone 1 (pre-vaporization, see Fig. 1), interphase thermal equilibrium is assumed, and hence they have the same temperature [24,26–29]. The Eulerian description is adopted for the liquid phase and hence the equation for droplet mass loading $Y_d$ ($\equiv \widetilde{N}_d \widetilde{m}_d / \tilde{\rho}_g$) reads

$$\frac{\partial}{\partial \tilde{t}}\left(\frac{\widetilde{N}_d \widetilde{m}_d}{\tilde{\rho}_g}\right) = \frac{\partial Y_d}{\partial \tilde{t}} = -\frac{\widetilde{\omega}_v}{\tilde{\rho}_g}, \quad (5)$$

where $\widetilde{N}_d$ is the droplet number density.

We assume that the heat transferred from the surrounding gas to the droplets is completely used for phase change, which is related to the latent heat of evaporation $\tilde{q}_v$ [4,26,27,29]. Therefore, $\widetilde{\omega}_v$ in Eqs. (1), (2), (4) and (5) can be modelled as

$$\widetilde{\omega}_v = \frac{\widetilde{N}_d \tilde{s}_d \tilde{h}(\tilde{T}-\tilde{T}_v) H(\tilde{T}-\tilde{T}_v)}{\tilde{q}_v}, \quad (6)$$

where $\tilde{s}_d = \pi \tilde{d}^2$ is the surface area of a single droplet, $\tilde{d}$ is the droplet diameter, $Nu$ is the Nusselt number, $\tilde{T}_v$ is the boiling temperature of the liquid fuel. $\tilde{h}$ is the heat transfer coefficient, estimated using the Ranz and Marshall correlation [37]

$$Nu = \frac{\tilde{h}\tilde{d}_p}{\tilde{\lambda}_g} = 2.0 + 0.6\, Re^{1/2}\, Pr^{1/3}, \quad (7)$$

where $Nu$, $Pr$ and $Re$ are the Nusselt number, Prandtl number and particle Reynolds number, respectively. We can neglect the effect of particle Reynolds number due to the assumption of kinematic equilibrium and therefore $Nu \approx 2$. Accordingly, the evaporation rate $\widetilde{\omega}_v$ can be re-written as

$$\widetilde{\omega}_v = \widetilde{N}_d \tilde{s}_d \tilde{\lambda}_g Nu(\widetilde{T} - \widetilde{T}_v) H(\widetilde{T} - \widetilde{T}_v)/(\check{d}\check{q}_v). \tag{8}$$

To render the analytical analysis more general, normalization of Eqs. (1), (2) and (5) can be performed, with the following non-dimensional parameters

$$U = \frac{\tilde{u}}{\tilde{u}_b}, r = \frac{\tilde{r}}{\tilde{l}_{th}}, t = \frac{\tilde{t}}{\frac{\tilde{l}_{th}}{\tilde{u}_b}}, Y = \frac{\widetilde{Y}}{\widetilde{Y}_0}, T = \frac{\widetilde{T} - \widetilde{T}_0}{\widetilde{T}_b - \widetilde{T}_0}. \tag{9}$$

Here $\widetilde{T}_0$ and $\widetilde{Y}_0$ denote the temperature and fuel mass fraction of the pre-vaporized pre-mixture, respectively. $\tilde{u}_b$, $\widetilde{T}_b = \widetilde{T}_0 + \tilde{q}_c \widetilde{Y}_0/\widetilde{C}_{p,g}$ and $\tilde{l}_{th} = \widetilde{D}_{th}/\tilde{u}_b$ are respectively the laminar flame speed, adiabatic flame temperature and flame thickness based on the pre-vaporized pre-mixture. $\widetilde{D}_{th} = \tilde{\lambda}_g/\tilde{\rho}_g \widetilde{C}_{p,g}$ is the thermal diffusivity.

Following previous theoretical analysis for both gaseous flames and two-phase flames with dispersed liquid droplets [24,28,32,38–43], we adopt the quasi-steady state assumption in the moving coordinate system attached to the stably propagating flame front $R_f(t)$, i.e. $\eta = r - R_f(t)$. This assumption has been extensively validated by transient numerical simulations for gaseous spherical flames [32,38–40,44], in which the unsteady effects are found to have a negligible influence based on the budget analysis of diffusion, reaction and convection terms in stably propagating spherical flames. Moreover, due to relatively dilute fuel droplet concentration, their influences on the reaction zone thickness are small and therefore gaseous combustion still dominates [24,27]. In addition, due to the kinematic equilibrium between the two phases, the droplets approximately follow the carrier gas. Therefore, the non-dimensional Eqs. (1), (2) and (5) can be written as

$$-U\frac{dT}{d\eta} = \frac{1}{(\eta+R_f)^2}\frac{d}{d\eta}\left[(\eta+R_f)^2\frac{dT}{d\eta}\right] + \omega_c - q_v\omega_v + \alpha\bar{\omega}, \tag{10}$$

$$-U\frac{dY_F}{d\eta} = \frac{Le^{-1}}{(\eta+R_f)^2}\frac{d}{d\eta}\left[(\eta+R_f)^2\frac{dY_F}{d\eta}\right] - \omega_c + \omega_v - \alpha\bar{\omega}, \tag{11}$$

$$-U\frac{dY_d}{d\eta} = -\omega_v, \tag{12}$$

where $U = dR_f/dt$ is the non-dimensional flame propagating speed. $q_v = \tilde{q}_v/[\tilde{C}_{p,g}(\tilde{T}_b - \tilde{T}_0)]$ is the normalized latent heat of vaporization. $Le = \tilde{D}_{th}/\tilde{D}_F$ is the Lewis number. The normalized chemical reaction rate $\omega_c$ reads

$$\omega_c = \frac{1}{2Le}Y_F Z^2 \exp\left[\frac{Z(T-1)}{\sigma+(1-\sigma)T}\right], \tag{13}$$

where $Z$ is the Zel'dovich number and $\sigma$ is the thermal expansion ratio. They are assumed to be $Z = 10$ and $\sigma = 0.15$, respectively, following Refs. [24,26,28,29,32,33].

The term $\omega_v$ in Eqs. (10)−(12) is the non-dimensional droplet evaporation rate, i.e.

$$\omega_v = \frac{\Omega(T-T_v)}{q_v}H(T-T_v), \tag{14}$$

where $T_v$ is the non-dimensional boiling temperature and assumed to be $T_v = 0.15$ [24,45]. The heat exchange coefficient $\Omega$ is

$$\Omega = \pi\tilde{N}_d Nu\tilde{d}\tilde{D}_{th}^2\tilde{u}_b^{-2}. \tag{15}$$

For a fixed latent heat $q_v$, higher $\Omega$ indicates faster droplet evaporation rate. Moreover, the normalized droplet burning term $\bar{\omega}$ in Eqs. (10) and (11) is

$$\bar{\omega} = \omega_v H(-\eta). \tag{16}$$

In the current work, propagation of spherical spray flames under moderate or weak stretch rate will be investigated. Therefore, we assume that the reactive−diffusive structure of the spherical flame is quasi-planar ($R_f \gg 1$, $\eta \sim O(1)$) [26,40,42,43,46] in this study. Its validity has been confirmed in Refs. [40,46]. Therefore, the governing equations of gas and liquid phases are reduced to

$$\frac{d^2 T}{d\eta^2} + \left(\frac{2}{R_f} + U\right)\frac{dT}{d\eta} + \omega_c - q_v \omega_v + \alpha \bar{\omega} = 0, \tag{17}$$

$$\frac{d^2 Y_F}{d\eta^2} + \left(\frac{2}{R_f} + LeU\right)\frac{dY_F}{d\eta} - Le\omega_c + Le\omega_v - \alpha Le\bar{\omega} = 0, \tag{18}$$

$$-U\frac{dY_d}{d\eta} = -\omega_v. \tag{19}$$

### 2.3   Jump and boundary conditions

The non-dimensional boundary conditions for both gas phase ($T$ and $Y_F$) and liquid phase ($Y_d$) at the left boundary (spherical center, $\eta = -R_f$) and the right boundary ($\eta \to +\infty$) are [24,26–29]

$$\eta = -R_f: \quad \frac{dT}{d\eta} = 0, \; \frac{dY_F}{d\eta} = 0, \; Y_d = \begin{cases} 0, & if\ \eta_c \neq -R_f \\ \delta_r, & if\ \eta_c = -R_f \end{cases}, \tag{20}$$

$$\eta \to +\infty: \quad T = 0,\; Y_F = 1,\; Y_d = \delta. \tag{21}$$

Here $\delta$ is the initial mass loading of the fuel droplet in the fresh mixture. In Eq. (20), for $Y_d$, if $\eta_c \neq -R_f$, the droplets at the spherical center are fully vaporized and hence partially distributed behind the flame. Otherwise, they exist in the entire burned area corresponding to $\eta_c = -R_f$, and $\delta_r$ is the mass loading at the spherical center to be determined (will be discussed later).

At the evaporation onset front, $\eta = \eta_v$, the gas temperature ($T$), fuel mass fraction ($Y_F$), and fuel droplet mass loading ($Y_d$) satisfy the following jump conditions [24,26–29]

$$T = T_v, \ [T] = [Y_F] = \left[\frac{dY_F}{d\eta}\right] = 0, Y_d = \delta. \tag{22}$$

where the square brackets, i.e. $[T] = T(\eta^+) - T(\eta^-)$, denote the difference between the variables at two sides of a location.

At the evaporation completion front, $\eta = \eta_c$, the jump conditions for the gas temperature ($T$), fuel mass fraction ($Y_F$), and droplet mass loading ($Y_d$) take the following form [27]

$$\begin{cases} [T] = [Y_F] = \left[\frac{dY_F}{d\eta}\right] = 0, Y_d = 0, & \eta_c > 0 \\ [T] = 0, \frac{dT}{d\eta}\Big|_+ = 0, [Y_d] = 0, & -R_f < \eta_c < 0 \end{cases}, \tag{23}$$

Here, the + symbol indicate the value is on the positive side of $\eta_c$, i.e. the left boundary of zone 3 in Fig. 1(a).

Large activation energy of the gas phase reaction is assumed in this study. This assumption has been successfully used for theoretical analysis of both gaseous [32,38,39,44,47,48] and particle- or droplet-laden [23,24,29,33–36] flames. It has been shown to be adequate to predict the main flame dynamics, such as ignition and propagation. In the limit of large activation energy, chemical reaction is confined at an infinitesimally thin sheet (i.e. $\eta = 0$). The corresponding jump conditions are

$$T = T_f, \ Y_F = [Y_d] = 0, \tag{24}$$

$$-\left[\frac{dT}{d\eta}\right] = \frac{1}{Le}\left[\frac{dY_F}{d\eta}\right] = \left[\sigma + (1-\sigma)T_f\right]^2 \exp\left[\frac{Z}{2}\left(\frac{T_f-1}{\sigma+(1-\sigma)T_f}\right)\right], \quad (25)$$

where $T_f$ is the flame temperature.

It should be highlighted that there exist two critical cases which only involves three zones. They respectively correspond to the coincidence of the evaporation completion front with spherical center ($\eta_c = -R_f$) and flame front ($\eta_c = 0$). The first is possible with the large droplet diameter, high initial liquid loading or low evaporation rate, leading to droplet dispersion in the full post-flame zone [13,14]. Accordingly, zone 4 (gaseous post-flame zone in Fig. 1a) is degenerated. The condition of $Y_d = \delta_r$ (see Eq. 20) is enforced at the spherical center, and $\delta_r$ is solved as an eigenvalue of the problem, instead of $\eta_c$. Moreover, the second case, i.e. $\eta_c = 0$, corresponds to a critical condition between heterogeneous and homogeneous flames. Zone 3 is degenerated, which is pre-flame (post-flame) evaporation zone in homogeneous (heterogeneous) flames. This is experimentally observed by Sulaiman et al. [15] and de Oliveira and Mastorakos [13], in which well-sprayed fuel droplets are fully vaporized around the flame. For this case, the jump conditions at the flame front (Eqs. 24 and 25) are used, instead of those at $\eta = \eta_c$ given in Eq. (23). This limiting situation is also studied by Zhuang and Zhang [29] for spherical flame propagation in fine water mists.

## 3 Analytical solution

Equations (17) – (19) with boundary and jump conditions (i.e. Eqs. 20 – 25) can be solved analytically. The solutions for gas temperature $T$, fuel mass fraction $Y_F$, and droplet mass loading $Y_d$ in four zones are presented in Section 3.1 for general heterogeneous and homogeneous flames, i.e. $\eta_c \neq -R_f$ or $0$. Moreover, the correlations for flame speed $U$, flame temperature $T_f$, evaporation onset location $\eta_v$ and completion location $\eta_c$ under different flame radii $R_f$ are derived in Section 3.2.

## 3.1 Distributions of $T$, $Y_F$ and $Y_d$

If the reactants around the flame front $R_f$ are heterogeneous (including fuel vapors and droplets), the distributions of temperature $T$, fuel mass fraction $Y_F$, and droplet loading $Y_d$ from zones 1 to 4 are (the number subscripts indicate different zones)

$$\begin{cases} T_1(\eta) = T_v e^{-\xi(\eta-\eta_v)} \\ T_2(\eta) = T_v - \frac{T_v \xi \left[e^{\gamma_a(\eta-\eta_v)} - e^{\gamma_b(\eta-\eta_v)}\right]}{\gamma_a - \gamma_b} \\ T_3(\eta) = T_v + \frac{T_f - T_v}{\mu_a - \mu_b} \left[\frac{e^{\chi_a(\eta-\eta_c)}}{\chi_a} - \frac{e^{\chi_b(\eta-\eta_c)}}{\chi_b}\right] \\ T_4(\eta) = T_v + \frac{T_f - T_v}{\mu_a - \mu_b} \left(\frac{1}{\chi_a} - \frac{1}{\chi_b}\right) \end{cases} \quad (26)$$

$$\begin{cases} Y_{F,1}(\eta) = 1 + \frac{I_2'(\eta_v)}{I_1'(\eta_v)} I_1(\eta) + G(\eta_v) I_1(\eta) \\ Y_{F,2}(\eta) = 1 + \frac{I_2'(\eta_v)}{I_1'(\eta_v)} I_1(\eta_v) + G(\eta_v) I_1(\eta) + I_2(\eta) - I_2(\eta_v) \\ Y_{F,3}(\eta) = -\frac{I_3'(\eta_c)}{I_1'(\eta_c)} I_1(\eta) + \frac{I_3'(\eta_c)}{I_1'(\eta_c)} I_1(0) + I_3(\eta) - I_3(0) \\ Y_{F,4}(\eta) = -\frac{I_3'(\eta_c)}{I_1'(\eta_c)} I_1(\eta_c) + \frac{I_3'(\eta_c)}{I_1'(\eta_c)} I_1(0) + I_3(\eta_c) - I_3(0) \end{cases} \quad (27)$$

$$\begin{cases} Y_{d,1}(\eta) = \delta \\ Y_{d,2}(\eta) = \delta - \frac{\Omega T_v \xi R_f}{U q_v (\gamma_a - \gamma_b)} \left[\frac{e^{\gamma_a(\eta-\eta_v)} - 1}{\gamma_a} - \frac{e^{\gamma_b(\eta-\eta_v)} - 1}{\gamma_b}\right] \\ Y_{d,3}(\eta) = \frac{\Omega(T_f - T_v)}{U q_v (\mu_a - \mu_b)} \left[\frac{e^{\chi_a(\eta-\eta_c)} - 1}{\chi_a^2} - \frac{e^{\chi_b(\eta-\eta_c)} - 1}{\chi_b^2}\right] \\ Y_{d,4}(\eta) = 0 \end{cases} \quad (28)$$

where $\xi = (2 + R_f U)/R_f$, $\gamma_{a,b} = \left(-\xi \pm \sqrt{4\Omega + \xi^2}\right)/2$, $\chi_{a,b} = \left(-\xi \pm \sqrt{4\Omega(1 - \alpha/q_v) + \xi^2}\right)/2$, and $\mu_{a,b} = e^{-\chi_{a,b}\eta_c}/\gamma_{a,b}$. $I_1(\eta)$, $I_2(\eta)$ and $I_3(\eta)$ are

$$I_1(\eta) = -\frac{Re^{-\frac{2+LeR_f U}{R}\eta}}{2+LeR_f U}, \tag{29}$$

$$I_2(\eta) = \frac{LeT_v(2+R_f U)\Omega}{q_v(\gamma_a-\gamma_b)} \left[\frac{e^{\gamma_a(\eta-\eta_v)}}{\gamma_a(2+\gamma_a R_f+LeR_f U)} - \frac{e^{\gamma_b(\eta-\eta_v)}}{\gamma_b(2+\gamma_b R_f+LeR_f U)}\right], \tag{30}$$

$$I_3(\eta) = \frac{LeR_f\Omega(T_f-T_v)(\alpha-1)}{q_v(\mu_a-\mu_b)} \left[\frac{e^{\chi_a(\eta-\eta_c)}}{\chi_a^2(2+\chi_a R_f+LeR_f U)} - \frac{e^{\chi_b(\eta-\eta_c)}}{\chi_b^2(2+\chi_b R_f+LeR_f U)}\right]. \tag{31}$$

$I_1'(\eta)$, $I_2'(\eta)$ and $I_3'(\eta)$ are respectively their first derivatives. In Eq. (27), $G(\eta_v)$ takes the following form

$$G(\eta_v) = \frac{-1-\frac{I_2'(\eta_v)}{I_1'(\eta_v)}I_1(\eta_v)+I_2(\eta_v)-I_2(0)}{I_1(0)}. \tag{32}$$

If the reactants around the flame front $R_f$ are homogeneous (i.e. fuel vapors only), the distributions of $T$, $Y_F$, and $Y_d$ from zones 1 to 4 are

$$\begin{cases} T_1(\eta) = T_v e^{-\xi(\eta-\eta_v)} \\ T_2(\eta) = T_v - \frac{T_v\xi[e^{\gamma_a(\eta-\eta_v)}-e^{\gamma_b(\eta-\eta_v)}]}{\gamma_a-\gamma_b} \\ T_3(\eta) = T_v - \frac{T_v\xi(\theta_a-\theta_b)}{\gamma_a-\gamma_b} - \frac{T_v(\gamma_a\theta_a-\gamma_b\theta_b)}{\gamma_a-\gamma_b}\left[1-e^{-\xi(\eta-\eta_c)}\right] \\ T_4(\eta) = T_f \end{cases}, \tag{33}$$

$$\begin{cases} Y_{F,1}(\eta) = 1 + \frac{I_2'(\eta_v)}{I_1'(\eta_v)}I_1(\eta) - \mathcal{H}(\eta_v,\eta_c)I_1(\eta) \\ Y_{F,2}(\eta) = 1 + \frac{I_2'(\eta_v)}{I_1'(\eta_v)}I_1(\eta_v) - \mathcal{H}(\eta_v,\eta_c)I_1(\eta) + I_2(\eta) - I_2(\eta_v) \\ Y_{F,3}(\eta) = \left[\mathcal{H}(\eta_v e\eta_c) - \frac{I_2'(\eta_c)}{I_1'(\eta_c)}\right][I_1(0) - I_1(\eta)] \\ Y_{F,4}(\eta) = 0 \end{cases}, \tag{34}$$

$$\begin{cases} Y_{d,1}(\eta) = \delta \\ Y_{d,2}(\eta) = \delta - \frac{\Omega T_v \xi R_f}{U q_v (\gamma_a - \gamma_b)} \left[ \frac{e^{\gamma_a(\eta - \eta_v)} - 1}{\gamma_a} - \frac{e^{\gamma_b(\eta - \eta_v)} - 1}{\gamma_b} \right], \\ Y_{d,3}(\eta) = 0 \\ Y_{d,4}(\eta) = 0 \end{cases} \qquad (35)$$

where $\theta_{a,b} = e^{\gamma_{a,b}(\eta_c - \eta_v)}$. In Eq. (34), $\mathcal{H}(\eta_v, \eta_c)$ has the following form

$$\mathcal{H}(\eta_v, \eta_c) = \frac{1 + \frac{I_2'(\eta_v)}{I_1'(\eta_v)} I_1(\eta_v) + \frac{I_2'(\eta_c)}{I_1'(\eta_c)} [I_1(0) - I_1(\eta_c)] - I_2(\eta_v) + I_2(\eta_c)}{I_1(0)}. \qquad (36)$$

## 3.2   Correlations for spherical flame and fuel sprays

If the reactants around the flame front $R_f$ are heterogeneous, then the following correlation holds

$$\frac{T_f - T_v}{\mu_a - \mu_b} \left( e^{-\chi_a \eta_c} - e^{-\chi_b \eta_c} \right) + \frac{\xi T_v}{\gamma_a - \gamma_b} \left( \gamma_a e^{-\gamma_a \eta_v} - \gamma_b e^{-\gamma_b \eta_v} \right) = Q(T_f), \qquad (37)$$

$$\frac{G(\eta_v) I_1'(0) + I_2'(0) + \frac{I_3'(\eta_c)}{I_1'(\eta_c)} I_1'(0) - I_3'(0)}{Le} = Q(T_f), \qquad (38)$$

$$T_v + \frac{\xi T_v}{\gamma_b - \gamma_a} \left( e^{-\gamma_a \eta_v} - e^{-\gamma_b \eta_v} \right) = T_f, \qquad (39)$$

$$\delta - \frac{\Omega T_v \xi R_f}{U q_v (\gamma_a - \gamma_b)} \left( \frac{e^{-\gamma_a \eta_v} - 1}{\gamma_a} - \frac{e^{-\gamma_b \eta_v} - 1}{\gamma_b} \right) = \frac{\Omega (T_f - T_v)}{U q_v (\mu_a - \mu_b)} \left( \frac{e^{-\chi_a \eta_c} - 1}{\chi_a^2} - \frac{e^{-\chi_b \eta_c} - 1}{\chi_b^2} \right), \qquad (40)$$

where $Q(T_f)$ is the normalized chemical heat release at the flame front

$$Q(T_f) = \left[ \sigma + (1 - \sigma) T_f \right]^2 \exp \left[ \frac{Z}{2} \left( \frac{T_f - 1}{\sigma + (1 - \sigma) T_f} \right) \right]. \qquad (41)$$

If the reactants around the flame front $R_f$ are homogeneous, then the correlation is

$$\frac{\xi T_v}{\gamma_a - \gamma_b}[\gamma_a \theta_a - \gamma_b \theta_b] e^{\xi \eta_c} = Q(T_f), \tag{42}$$

$$-I_1'(0)\left[\mathcal{H}(\eta_v, \eta_c) - \frac{I_2'(\eta_c)}{I_1'(\eta_c)}\right]/Le = Q(T_f), \tag{43}$$

$$T_v + \frac{T_v}{\gamma_a - \gamma_b}(\gamma_a \theta_a - \gamma_b \theta_b)(1 - e^{\xi \eta_c}) - \frac{\xi T_v}{\gamma_a - \gamma_b}(\theta_a - \theta_b) = T_f, \tag{44}$$

$$\frac{\Omega \xi T_v}{U q_v (\gamma_b - \gamma_a)}\left[\frac{1 - e^{\gamma_a(\eta_c - \eta_v)}}{\gamma_a} - \frac{1 - e^{\gamma_b(\eta_c - \eta_v)}}{\gamma_b}\right] = \delta. \tag{45}$$

Furthermore, the analytical solutions and correlations corresponding to the two critical scenarios mentioned in Section 2.3, i.e. $\eta_c = 0$ and $\eta_c = -R_f$, will be provided in the Supplemental Material. Besides, the results for homogeneous flame from the current model can recover the solutions for gaseous flame [32] in the limits of $\delta \to 0$, $\Omega \to 0$ and $R_f \to +\infty$, which are also presented in the Supplemental Material.

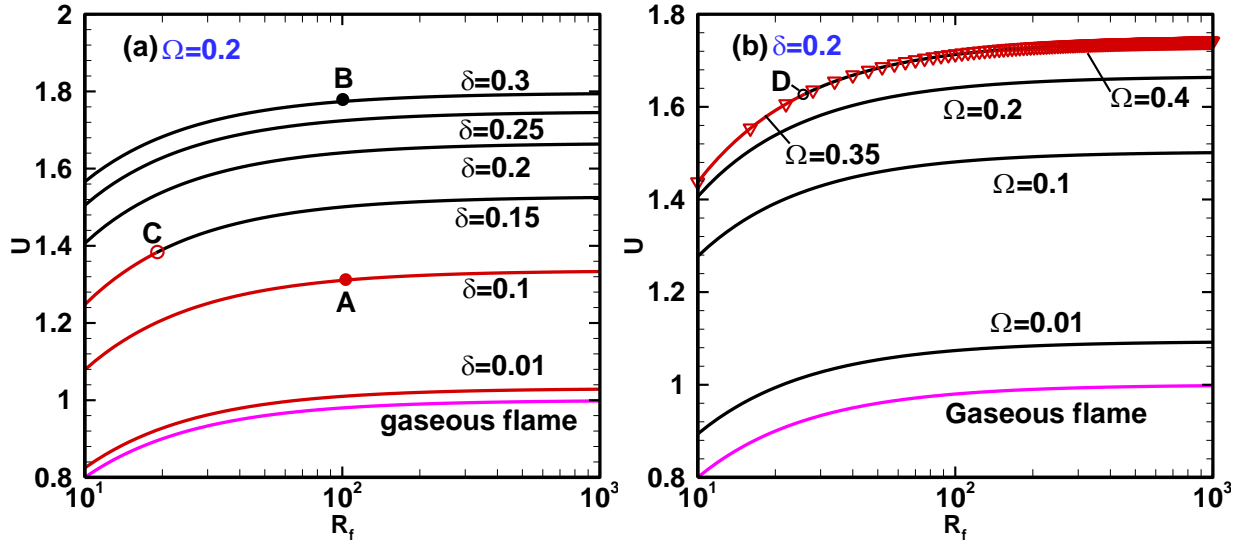

Fig. 2. Changes of flame propagation speed with flame radius for different (a) droplet mass loadings and (b) evaporative heat exchange coefficients. $Le = 1$, $q_v = 0.4$. Red line and triangles: homogeneous; black line: heterogeneous.

## 4  Results and discussion

With the correlations in Section 3.2, spherical flame propagation in pre-vaporized liquid fuel sprays will be studied. In Sections 4.1–4.5, both heterogeneous (without droplet burning, $\alpha = 0$) and homogeneous flames will be discussed. In Section 4.6, the influence of droplet burning behind the heterogeneous flame ($\alpha = 1$) will be studied.

### 4.1  Spherical flame propagation

The effects of droplet properties (mass loading $\delta$, evaporative heat exchange coefficient $\Omega$, and latent heat of vaporization $q_v$) on spherical flame propagation will be first investigated. The Lewis number is fixed to be $Le = 1$. Figure 2(a) shows the change of flame propagation speed with various droplet mass loadings. Here $\Omega = 0.2$ and $q_v = 0.4$. It is seen that for all the mass loadings, the flame propagation speed rapidly increases with the flame radius due to enhanced diffusive flux, eventually approaching those of the un-stretched planar flames ($R_f \to +\infty$) [32]. Furthermore, for the same flame radius, propagation speed is higher than that of gaseous flame and increases with the initial droplet loading. This is because more liquid fuels are vaporized into the gas phase, rendering the composition closer to the stoichiometry. It is also found that when droplet loading is small (e.g. $\delta = 0.01$ and 0.1), the reactants around propagating spherical flame front are homogeneous. However, with $\delta \geq 0.2$, they are heterogeneous and composed of fuel vapor and droplets. This is because bigger $\delta$ corresponds to higher droplet concentration. Interestingly, for $\delta = 0.15$, the flame firstly propagates in homogeneous reactants, but at $R_f \approx 18.9$ (marked with circle C), the droplets in the unburned zone cross the moving flame front, and the spray flame transitions to heterogeneous combustion.

The propagation speed of spherical flames with various evaporative heat exchange coefficients $\Omega$ is presented in Fig. 2(b). Here $\delta = 0.2$ and $q_v = 0.4$. In general, the flame propagation speed is higher with larger $\Omega$ when the loading is fixed. This is justifiable since larger $\Omega$ means the faster droplet evaporation rate. However, this tendency becomes weak when $\Omega = 0.35$ and 0.4. For $\Omega = 0.35$ in Fig. 2(b), transition between homogeneous and heterogeneous flames at $R_f \approx 26$ (circle D) are observed, similar to the case with $\delta = 0.15$ in Fig. 2(a).

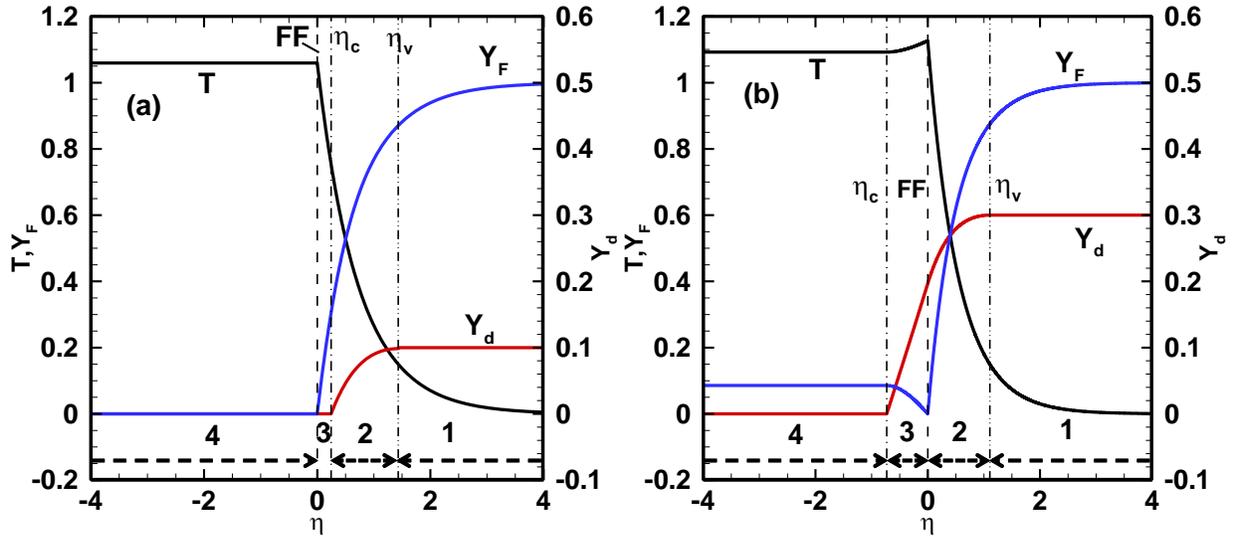

Fig. 3. Spatial distributions of temperature, fuel mass fraction and droplet mass loading for $R_f = 100$ when (a) $\delta = 0.1$ and (b) $\delta = 0.3$. Numbers indicate the flame and droplet zones. FF: flame front ($\eta = 0$).

Figures 3(a) and 3(b) show two flame structures with $\delta = 0.1$ and 0.3 (A and B in Fig. 2a), respectively. Their flame radii are $R_f = 100$. Note that Fig. 3(a) shows a homogeneous flame, while Fig. 3(b) a heterogeneous one. In Fig. 3(a), the droplets are fully vaporized (i.e. $Y_d \to 0$) slightly before the flame front ($\eta_c = 0.246$). The extra fuel addition from droplet evaporation increases the overall fuel concentration in the gaseous mixture, leading to higher temperature in the post-flame zone compared to that (= 1) of droplet-free flames. Meanwhile, due to no droplet evaporation in the post-flame zone ($\eta < 0$), the local temperature is uniform.

The heterogeneous flame structure in Fig. 3(b) is different from that of homogeneous one. Firstly, finite fuel concentration can be found in the post-flame zone, since evaporation of the penetrated droplets occurs and burning is not considered ($\alpha = 0$). Presence of fuel vapour behind the flame is also revealed by Greenberg and Kalma [49] from spherical flame with less volatile liquid fuel sprays. Part of them contribute towards the spray flame through back diffusion from the post-flame zone to the flame front, which is also observed from the simulations and experiments of propagating flame in fuel mists [13,19,50]. The post-flame evaporation and its evolving interaction with the leading stretch flame front are of high relevance to practical spray combustion, but has not been included in previous theoretical analysis [23,24]. Moreover, there are still residual fuel vapors beyond the evaporation zone in the burned area, i.e. $\eta < \eta_c$. This is different from the results in homogeneous flame, and the kinetic effects of heterogeneous flame arise from both pre- and post-flame evaporation zones (further interpreted in Section 4.3). Secondly, the gas temperature in the post-flame zone gradually decreases from the flame front in the post-flame evaporation zone. Heat conduction in this zone (i.e. $\eta_c < \eta < 0$) would also weaken the flame reactivity, due to the considerable temperature gradient near the flame front. This is also reported from the results of spherical premixed flame with water mists [26].

The latent heat of vaporization considered in Figs. 2 and 3 is $q_v = 0.4$. However, different liquid fuels may be used in practical combustion systems and they have different latent heat. Figure 4 shows flame propagation speed as a function of flame radius when $q_v = 1.2$. Four droplet mass loadings are considered in Fig. 4(a), i.e. $\delta = 0.01$, 0.1, 0.3, 0.5 and 0.7. The evaporative heat exchange coefficient $\Omega$ is assumed to be 0.0775. For $\delta \leq 0.5$, the single-valued flame speed monotonically increases when the spherical flame propagates outwardly. They are called as normal flames. Also, the flame propagation speed is consistently smaller than that of a gaseous flame. Meanwhile, for normal flames, the propagation speed decreases with droplet loading. Nevertheless, this trend is not pronounced when $\delta$ is beyond 0.3.

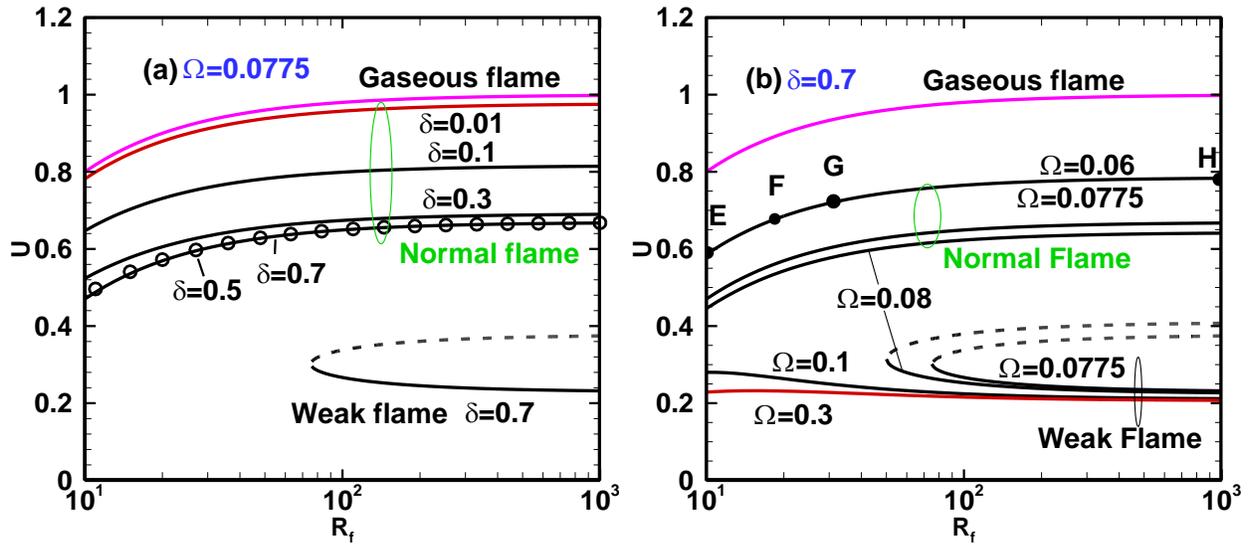

Fig. 4. Changes of flame propagation speed with flame radius for different (a) droplet mass loadings and (b) evaporative heat exchange coefficients. $Le = 1$, $q_v = 1.2$. Symbol and red line: homogeneous; black line: heterogeneous. Dotted line: $\eta_c = -R_f$.

For $\delta = 0.7$ in Fig. 4(a), flame bifurcation occurs. Besides the normal flame branch, a $C$-shaped branch is present when $R_f > 74$, with upper and lower solutions being unstable and stable flames. Therefore, three solutions are present when the flame radius exceeds a critical value. This is also observed in the previous studies on two-phase spherical flames laden with water mists [26,29]. These three flames respectively correspond to stable normal flame, unstable flame and stable weak flame. For normal flames, their behaviors of propagation speed, evaporation completion and onset fronts are similar to the results with smaller $\delta$. For weak flame, the propagation speed decreases with flame radius and is much lower than the normal flame.

The propagation speeds of spherical spray flame subject to various evaporation heat exchange coefficients are shown in Fig. 4(b). Here fixed $\delta = 0.7$ is considered. For small $\Omega$ (e.g. 0.06), the heat transfer due to droplet vaporization is slow, and the normal spray flames propagate outwardly. At some median values, i.e. $\Omega = 0.0775$ and 0.08, flame bifurcation occurs, similar to that in Fig. 4(a). Both normal and weak flames exist in Fig. 4(b) when $\Omega = 0.0775$ and 0.08. Meanwhile, increasing $\Omega$ would sustain the weak flame in a smaller flame radius. However, if $\Omega$ is even higher (0.1 or 0.3), the spray flame can only propagate in the weak flame mode due to fast droplet evaporation rate and strong heat absorption. Meanwhile, the large $\Omega$ would render the mixture heterogeneous around the flame front, which is also seen from Fig. 2(b).

One interesting phenomenon worth further discussion is that for $\Omega = 0.06$ in Fig. 4(b), the evaporation rate is so slow that the droplets are distributed in the full post-flame zone when the flame radius is small (line EF). At a critical flame radius ($R_f = 19.9$, point F), the droplets at the center ($r = 0$ or $\eta = -R_f$) are just fully vaporized. When the flame further grows, the droplet-free area centering at $r = 0$ expands and hence only part of the post-flame zone still has evaporating fuel droplets (line FH). To reveal this phenomenon, the flame structures at E, F and G are presented in Fig. 5. When $R_f = 10$ (point E), considerable gradients exist for temperature, fuel mass fraction and droplet loading in zone 3, as evaporating droplets are distributed in the entire burned area and the loading at the spherical center, $\delta_r$, is greater than zero (solid circle in Fig. 5a). This would lead to significant heat and mass transfer between the flame front and burned area. When $R_f = 19.9$ (point F), the droplets around the spherical center are just fully gasified. Thus, the droplet loading at $r = 0$ is $\delta_r = 0$ (circle in Fig. 5b). Here, the evaporation completion front lies at the spherical center, as the limiting case with $\eta_c = -R_f$, mentioned in Section 2.3. For these two scenarios, only three zones exist. For point G ($R_f = 30$), the droplets are dispersed in part of the post-flame zone. Thus, the evaporation completion front (circle in Fig. 5c) is deviated from the spherical center and moves concurrently with the leading flame front. Its structure is qualitatively like that in Fig. 3(b), despite the lower temperature in the post-flame zone due to larger latent heat. This corroborates the capacity of our theoretical model in predicting the fuel droplet dynamics in a propagating spray flame.

### 4.2 Fuel droplet distribution

The homogeneous and heterogeneous spherical spray flames discussed above are characterized by evolving droplet distributions when they propagate outwardly, which can be further quantified with evaporation onset and completion locations ($\eta_v$ and $\eta_c$). Figures 6(a) and 6(b) shows the evolutions of the two characteristic locations with various mass loadings $\delta$ and heat exchange coefficients $\Omega$, respectively. Their corresponding flame propagation speeds are shown in Fig. 2(a) and 2(b), respectively. Droplet evaporation always starts before the flame sweeps, parameterized by $\eta_v > 0$ in both figures. In Fig. 6(a), for fixed $\Omega$, when the droplet loading is increased, the droplets start to vaporize closer to the flame front. This is because larger $\delta$ leads to higher flame temperature and hence less spread temperature in the pre-flame zone [26].

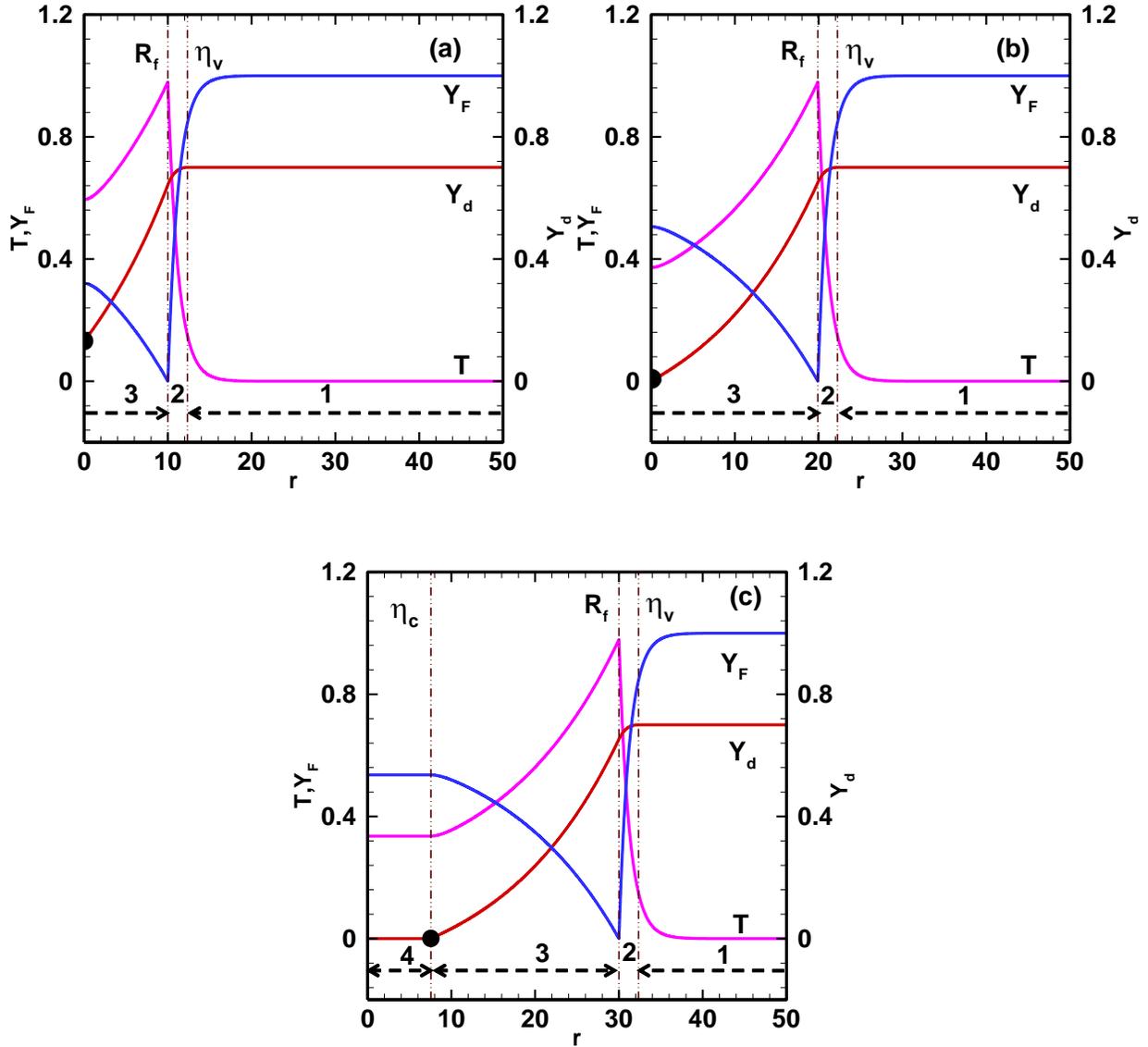

Fig. 5. Spatial distributions of temperature, fuel mass fraction and droplet mass loading for (a) $R_f = 10$, (b) $R_f = 19.9$, and (c) $R_f = 30$. They correspond to the circles E, F and G in Fig. 4, respectively. Numbers indicate the flame and droplet zones.

The variations of evaporation completion location $\eta_c$ are more complicated. When the loading $\delta$ is low (e.g. 0.01 and 0.1), the flame always propagates in a homogeneous mixture with fuel vapour only. Nevertheless, when $\delta$ is high (0.2−0.3), the reactants around the flame front are heterogeneous with fuel vapor and droplets. In particular, when $\delta$ is 0.15, the liquid droplets are completely vaporized when the flame radius is small, but slightly penetrate into the post-flame zone as the spherical flame expands. Therefore, evolution from homogeneous flame to heterogeneous flame occurs at point C in Fig. 6(a). This is reasonable since it takes longer time for the fuel droplet sprays to complete evaporation when δ is large. This phenomenon is also observed by de Oliveira and Mastorakos using simultaneous OH* chemiluminescence and OH-PLIF imaging of spray Jet A flames [13], and it is found that the local microscopic flame topology may be modulated by the encroaching droplets, which significantly affect the local mass and heat transfer near the flame front.

In Fig. 6(b), for fixed $\delta = 0.2$, change of $\Omega$ has limited effect on evaporation onset locations $\eta_v$. Moreover, higher $\Omega$ makes the evaporation completion front closer to the flame front ($\eta = 0$). Noted that for $\Omega = 0.35$, the flame experiences a transition from homogeneous to heterogeneous flame at $R_f \approx 26$ (marked with an open circle D). It is also found from Fig. 6(b) that the thickness of the droplet evaporation zone, i.e. $|\eta_c - \eta_v|$, increases with decreased $\Omega$. This is due to slower evaporation rate corresponding to smaller $\Omega$.

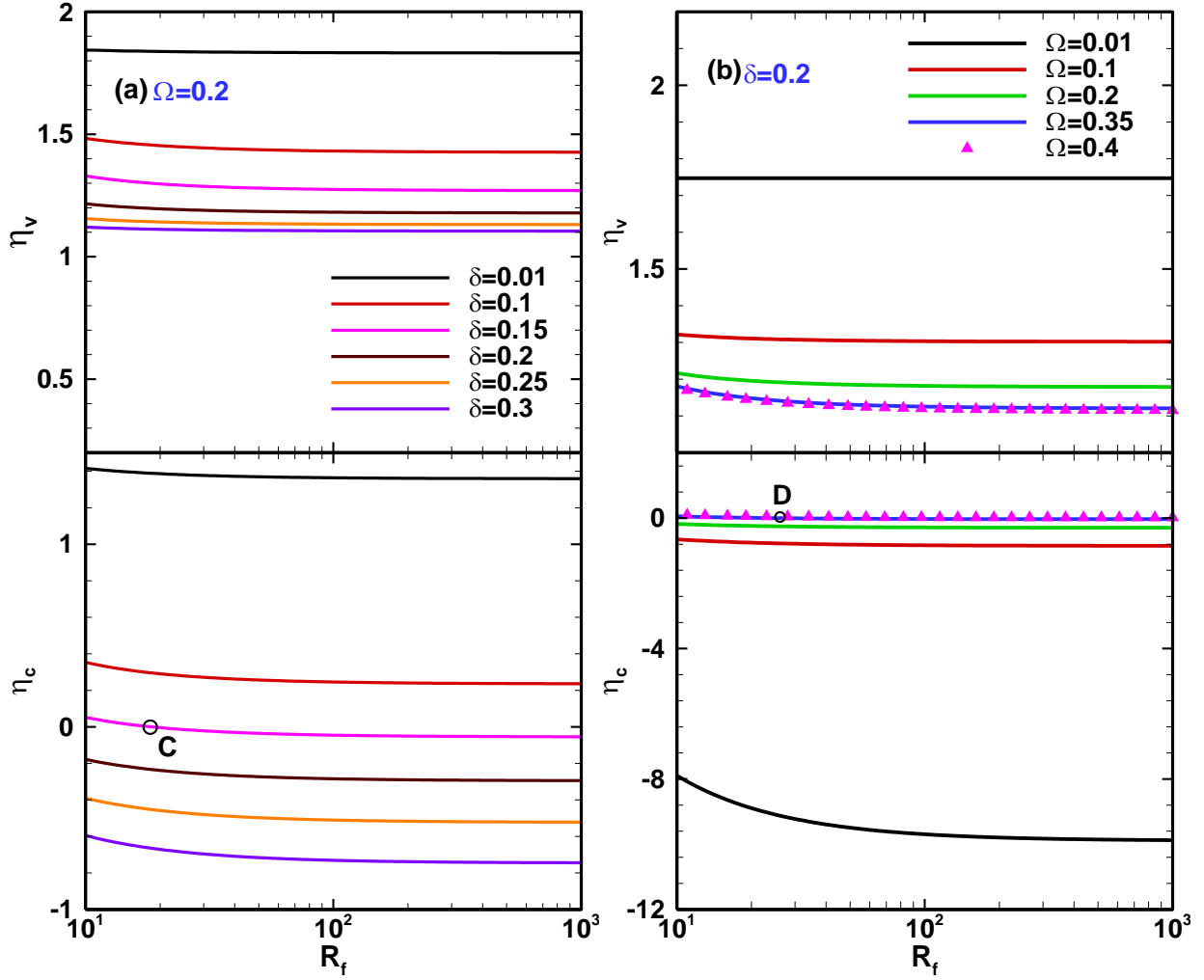

Fig. 6. Changes of evaporation onset and completion locations with flame radius for different (a) droplet mass loadings and (b) evaporative heat exchange coefficients. $q_v = 0.4$.

The various distributions of liquid droplets may have different thermal (evaporative cooling) and kinetic (fuel vapor addition) contributions towards the spherical spray flame. The evaporative cooling effects can be quantified by normalized evaporation heat loss $H$ [26,32]. In the unburned and burned zones of a heterogeneous flame, $H$ is respectively calculated as

$$H_{ub,HT} = \Omega \int_0^{\eta_v}[T_2(\xi) - T_v](\xi + R_f)^2 d\xi / \left(R_f^2 \frac{dT}{d\eta}\Big|_{-} - R_f^2 \frac{dT}{d\eta}\Big|_{+}\right), \qquad (46)$$

$$H_{b,HT} = \Omega \int_{\eta_c}^{0} [T_3(\xi) - T_v](\xi + R_f)^2 d\xi \Big/ \left(R_f^2 \frac{dT}{d\eta}\Big|_{-} - R_f^2 \frac{dT}{d\eta}\Big|_{+}\right). \tag{47}$$

For a homogeneous flame, they are

$$H_{ub,HM} = \Omega \int_{\eta_c}^{\eta_v} [T_2(\xi) - T_v](\xi + R_f)^2 d\xi \Big/ \left(R_f^2 \frac{dT}{d\eta}\Big|_{-} - R_f^2 \frac{dT}{d\eta}\Big|_{+}\right), \tag{48}$$

$$H_{b,HM} = 0. \tag{49}$$

The subscripts "$ub$" and "$b$" respectively denote evaporative heat loss from unburned and burned zones, whereas "HM" and "HT" respectively denote homogeneous and heterogeneous flames. The denominator, $\left(R_f^2 \frac{dT}{d\eta}\Big|_{-} - R_f^2 \frac{dT}{d\eta}\Big|_{+}\right)$, is the combustion heat release. Equation (49) is valid since there are no droplets in the burned zone of homogeneous flames. The total evaporation heat loss is $H_{all} = H_b + H_{ub}$.

The fuel addition from droplet evaporation can be parameterized by fuel vapour yield $f$ [24]. In the pre-flame evaporation zone, $f_{ub}$ is

$$f_{ub} = Y_d(\eta_v) - Y_d(0). \tag{50}$$

For a heterogeneous flame, in the post-flame zone, $f_b$ is

$$f_b = Y_d(0) - Y_d(\eta_c). \tag{51}$$

The total fuel vapour yield is $f_{all} = f_{ub} + f_b$. Note that for a homogeneous flame, $f_b$ is zero since no evaporation occurs behind the flame front. Accordingly, $f_{all} = f_{ub}$ holds.

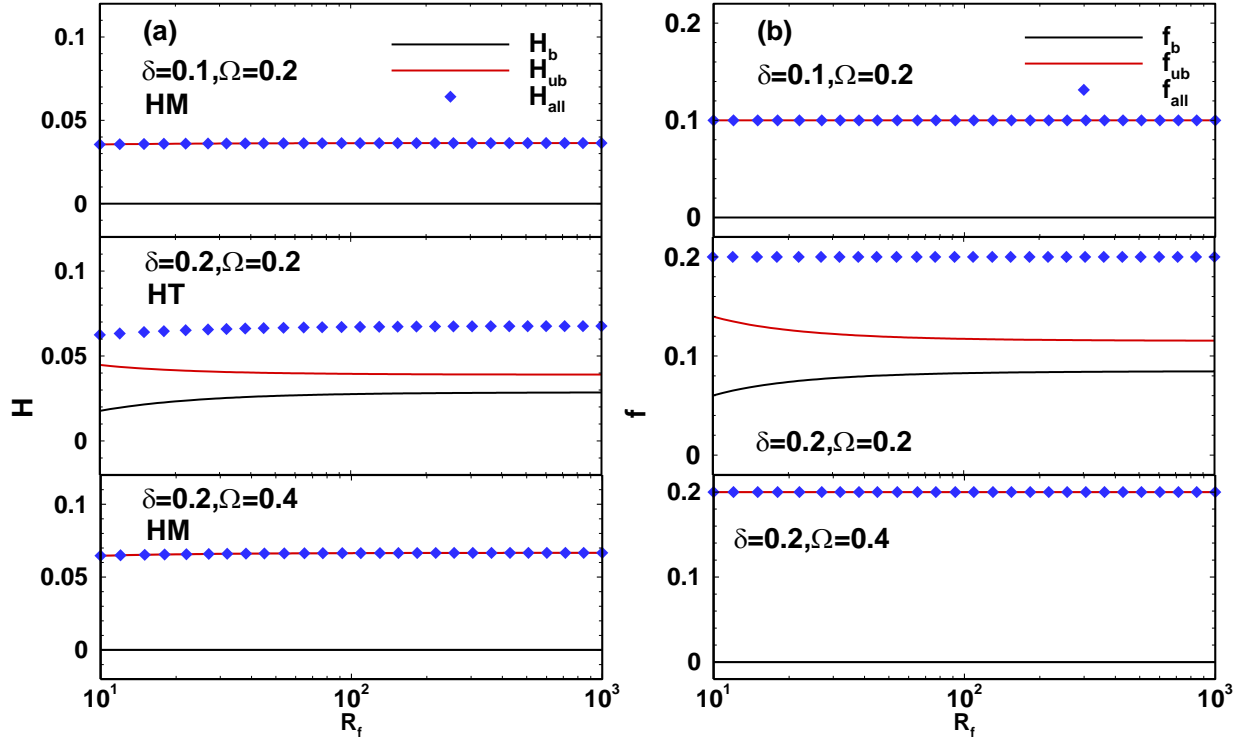

Fig. 7. Changes of (a) evaporation heat loss and (b) fuel vapor yield with flame radius.

Figure 7(a) shows that the normalized evaporative heat loss for three flames with different $\delta$ and $\Omega$ in Fig. 6. For the homogeneous flames, the evaporation induced heat transfer only occurs in the pre-flame zone and $H_{ub,HM}$ almost remains constant when the flame expands. Meanwhile, for the heterogeneous flame, $H_{ub,HT}$ ($H_{b,HT}$) decreases (increases) with the flame radius, which leads to monotonically (but subtly) increased $H_{all}$. Besides, the higher $\delta$, the higher $H_{all}$. Meanwhile, for fixed droplet loading, e.g. $\delta = 0.2$, $H_{all}$ is almost independent on $\Omega$.

Figure 7(b) shows the corresponding fuel vapour yield. For the homogeneous flames, the fuel vapor from droplet evaporation only exists in the pre-flame zone. For the heterogeneous flame, the fuel vapor exists in both pre- and post-flame zone. Note that the total fuel vapor yield $f_{all}$ is the same as the droplet loading $\delta$ for all the shown flames due to the complete droplet evaporation in the domain. With increased $\delta$ or decreased $\Omega$, $f_b$ increases. This is because more droplets cross the flame front and finish evaporation there.

The influences of liquid fuel latent heat of vaporization on droplet distribution in a propagating flame are examined in Fig. 8(a), which shows the evaporation onset and completion locations as functions of flame radius $R_f$. The latent heat $q_v$ is 1.2, higher than that in Figs. 6 and 7. The evaporative heat loss coefficient $\Omega$ is 0.0775, and the flame propagation speed has been studied in Fig. 4(a). For $\delta \leq 0.3$, the flame speed acceleration leads to more droplets behind the flame front, and the evaporation completion front penetrates further in the post-flame zone when the flame propagates outwardly. Meanwhile, the evaporation onset front deviates from the flame front when $\delta$ is increased. For $\delta = 0.7$, bifurcation of droplet evaporation characteristic locations is also observed, same as the flame bifurcation in Fig. 4(a). For the normal flame, the behaviors of evaporation completion and onset fronts are like those with smaller $\delta$. However, for the weak flame, the evaporation completion front moves closer to the flame front, since the droplet evaporation starts earlier when the flame front moves outwardly.

For fixed droplet loading $\delta = 0.7$ in Fig. 8(b), with decreased $\Omega$, the trends for both normal flame and weak flame are similar to those with increased $\delta$ in Fig. 8(a). However, for the normal flame with $\Omega = 0.06$, the fuel droplets are distributed in the entire post-flame zone at the early stage of flame propagation, which corresponds to the critical heterogeneous flames with $\eta_c = -R_f$ (triangles between E and F). This has been discussed in Figs. 4(b) and 5.

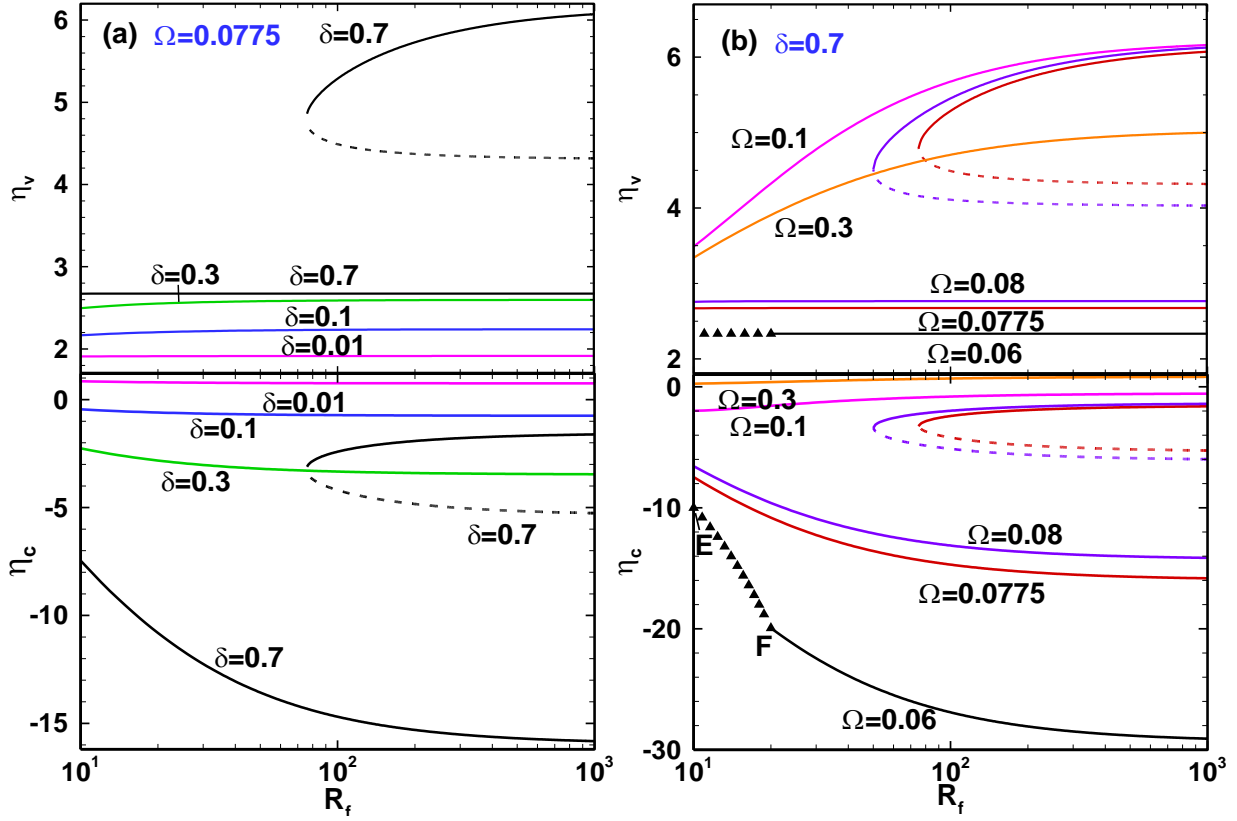

Fig. 8. Changes of evaporation onset and completion locations with flame radius for different (a) droplet mass loadings and (b) evaporative heat exchange coefficients. $Le = 1$ and $q_v = 1.2$. Dashed line: unstable solution; Triangle: $\eta_c = -R_f$.

Figure 9(a) shows that the normalized evaporative heat loss for the three flames selected from Fig. 8. It is found that when $\delta = 0.1$, variations of $H_{all}$ are limited when the flame propagates. Meanwhile, $H_b$ and $H_{ub}$ are comparable. When $\delta = 0.7$, for the normal flames with $\Omega = 0.06$ and 0.0775, $H_{all}$ and $H_b$ mainly contributes towards $H_{all}$, both of which gradually increase when the flame radius increases, while $H_{ub}$ is relatively small. This is because more droplets vaporize behind the flame front. For the weak flame with $\Omega = 0.0775$, $H_{all}$ does not change with the flame radius, with close contributions from $H_b$ and $H_{ub}$.

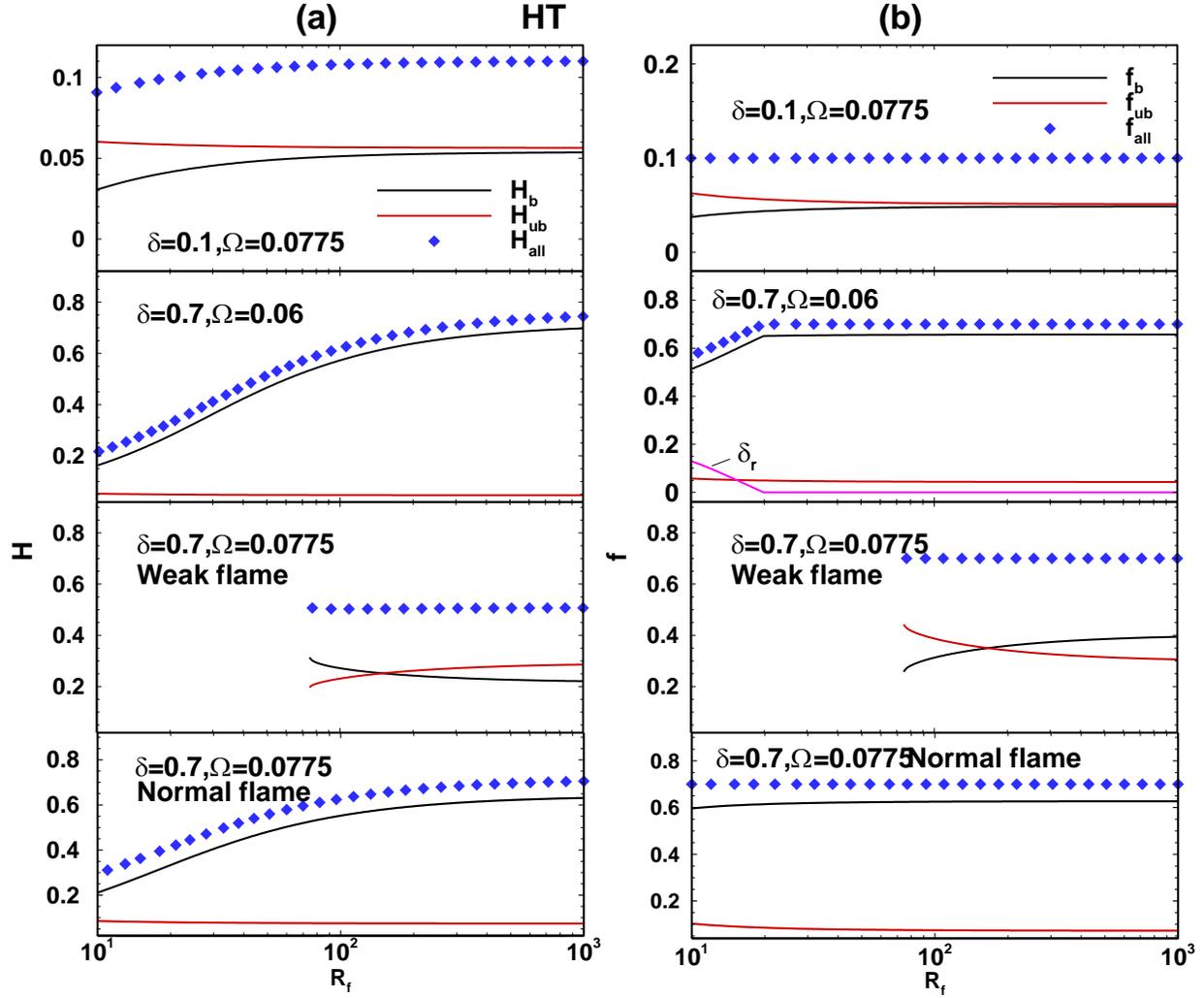

Fig. 9. Changes of (a) evaporation heat loss and (b) fuel vapor yield with flame radius.

Figure 9(b) shows the fuel vapor yield corresponding to the same heterogeneous flames in Fig. 9(a). The relations between $f_b$ and $f_{ub}$ are similar to those between $H_b$ and $H_{ub}$ indicated in Fig. 9(a). $f_{ub} = \delta$ holds if all the droplets are vaporized. However, for $\delta = 0.7$ and $\Omega = 0.06$, when the flame radius is small, droplets exist in the full domain, as mentioned in Figs. 5 and 8. Therefore, under this condition, $f_b$ and $f_{all}$ gradually increase with the droplet loading at the spherical center $\delta_r$ decreasing to zero (see Fig. 9b).

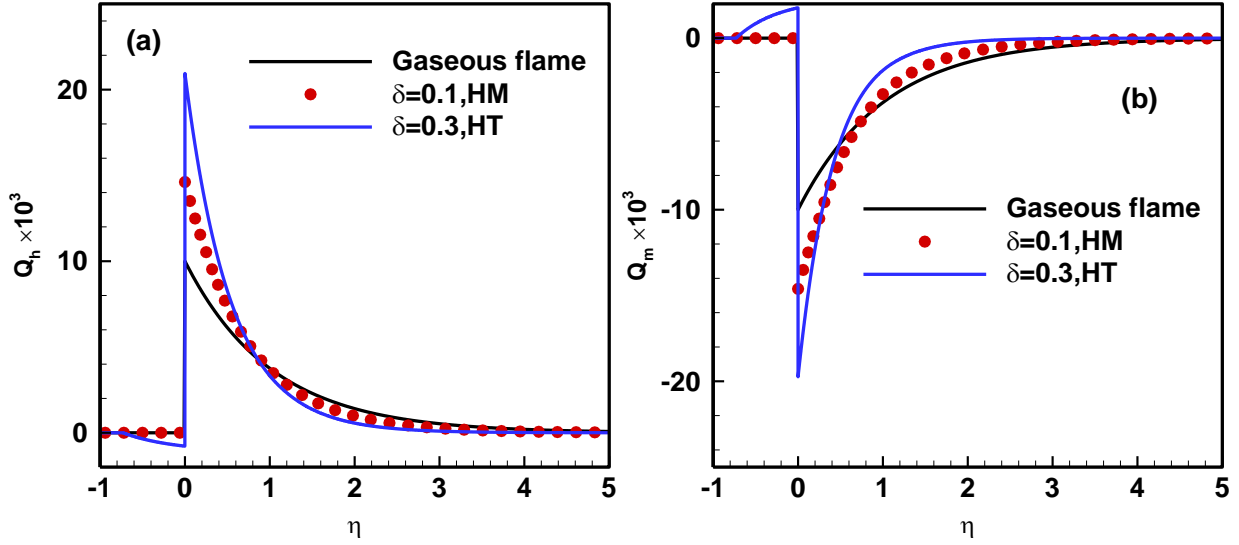

Fig. 10. Distributions of (a) heat flux and (b) mass flux. $R_f = 100$, $q_v = 0.4$ and $\Omega = 0.2$. The flame front lies at $\eta = 0$.

### 4.3 Diffusive fluxes in homogeneous and heterogeneous flames

To elucidate the respective influences of thermal and kinetic effects, diffusive heat and mass fluxes ($Q_h$ and $Q_m$) of homogeneous and heterogeneous flames are calculated as

$$Q_h = -(\eta + R_f)^2 \frac{dT}{d\eta}, \qquad (52)$$

$$Q_m = -\frac{(\eta + R_f)^2}{Le} \frac{dY_F}{d\eta}. \qquad (53)$$

Typical flames with Lewis number $Le = 1$ and flame radius $R_f = 100$ are selected for further analysis in Figs. 10 and 11. Figure 10 shows the heat and mass fluxes of homogeneous ($\delta = 0.1$) and heterogeneous flames ($\delta = 0.2$) with $q_v = 0.4$. They correspond to flames A and B in Fig. 2(a), respectively. Gaseous (no droplets, only pre-vaporized fuel) flame is also added for comparison. For gaseous and homogeneous spray flames, due to the uniform distributions of temperature and fuel mass fraction in the post-flame zone (see Fig. 3a), the heat and mass fluxes are zero. Conversely, heat conduction and fuel species diffusion only occur in the pre-flame zone.

For heterogeneous spray flame, the fluxes in the pre-flame zone show similar behaviours to homogeneous flame. However, due to the droplet evaporation in the post-flame zone, the fuel vapor is transported to the flame front, while heat released by the chemical reactions at the flame front is diffused into the post-flame evaporation zone to compensate the local evaporation heat loss. The magnitude of mass flux $|Q_m|$ is higher than that of heat flux $|Q_h|$, in spite of the same diffusion length. Moreover, both thermal and mass diffusion in homogeneous and heterogeneous flames are enhanced compared to the gaseous flame, evidenced by the higher flux magnitudes. This leads to stronger spray flames with higher propagation speeds. Also, near the flame front, $|Q_h|$ and $|Q_m|$ in heterogeneous flame are larger than the counterparts of homogeneous flames, due to stronger reactivity and hence higher gradient of $T$ and $Y_F$ in the former.

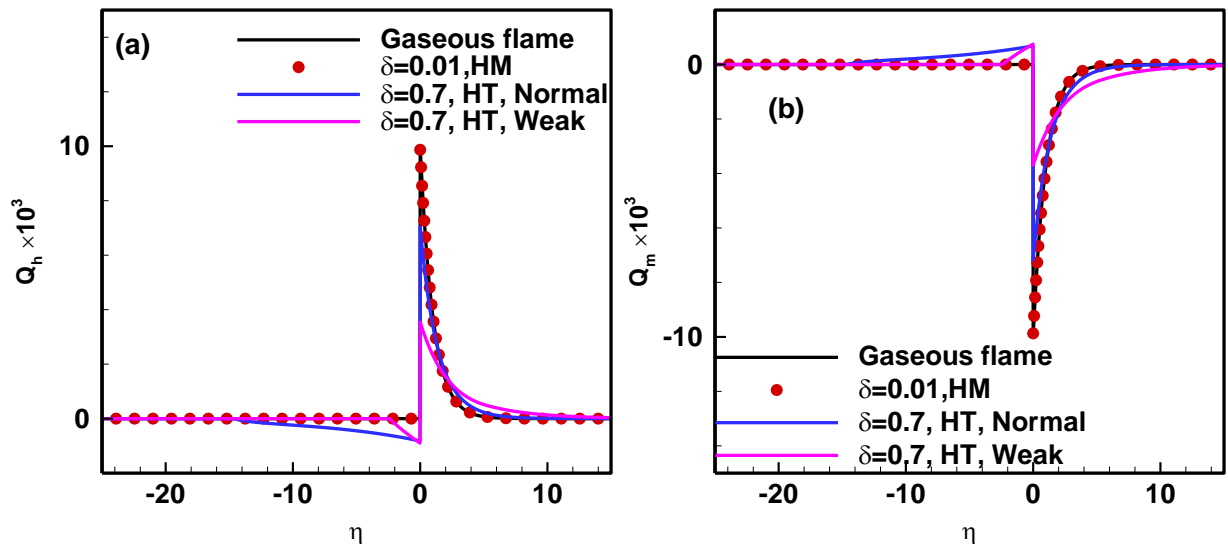

Fig. 11. Distributions of (a) heat flux and (b) mass flux. $R_f = 100$, $q_v = 1.2$ and $\Omega = 0.0775$. The flame front lies at $\eta = 0$.

The counterpart results with higher latent heat, i.e. $q_v = 1.2$, are presented in Fig. 11. They correspond to the flames with same parameters in Figs. 4(a) and 8(a). The trends for homogeneous and heterogeneous flames are similar to those in Fig. 10. For small $\delta$ like 0.01, the fuel droplet evaporation effect on the flame is limited, and both fluxes are similar to those of the gaseous flame. However, due to the weakening effect from larger $q_v$, with increased $\delta$, $|Q_h|$ and $|Q_m|$ decrease to a lower level compared to the gaseous flame. Furthermore, for $\delta = 0.7$, $|Q_h|$ and $|Q_m|$ of the weak flame are much smaller than those of the normal flame. Due to the low reactivity of the weak flame, its structure is more distributed with lower gradients of $T$ and $Y_F$. In the meanwhile, in the burned zones of the normal and weak flames, $|Q_m|$ is slightly larger than $|Q_h|$.

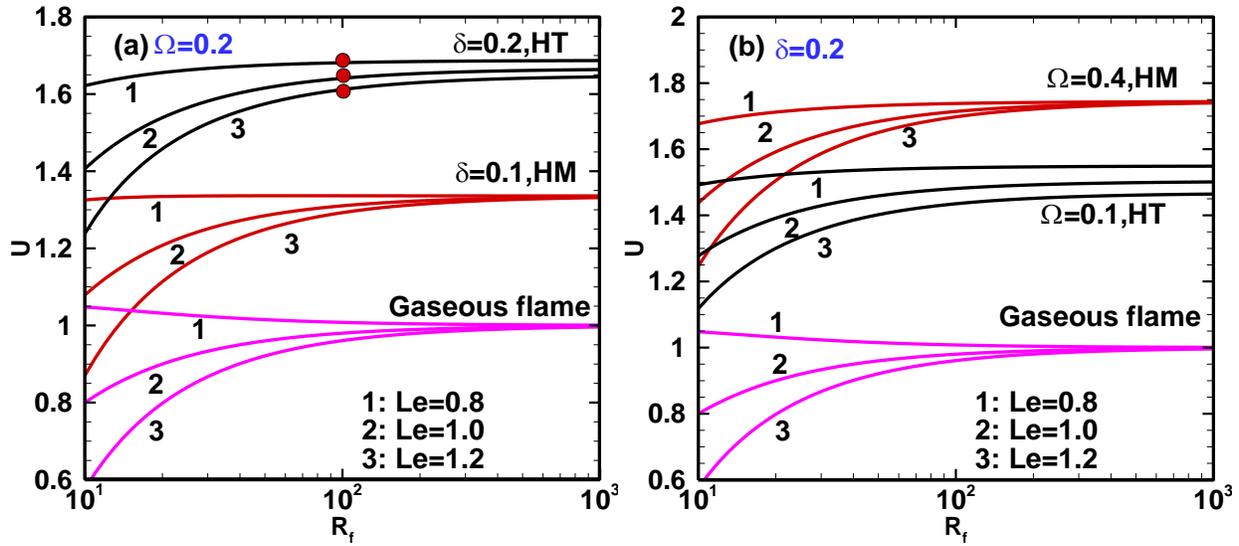

Fig. 12. Changes of flame propagation speed with flame radius for different Lewis numbers when (a) $\Omega = 0.2$ and (b) $\delta = 0.2$.

## 4.4 Lewis number effect

Up to this point, the Lewis number is assumed to be $Le = 1$. Figure 12(a) show the flame propagation speed $U$ as a function of flame radius $R_f$ when $Le = 0.8$, 1.0 and 1.2. Here $q_v = 0.4$ and $\Omega = 0.2$ are considered. It is seen that flame propagation speeds of both homogeneous and heterogeneous spherical flames are dependent on Lewis number. Generally, with each Lewis number, $U$ increases with $\delta$, because of the spray enrichment effects [51]. Furthermore, $U$ decreases with Lewis number when $R_f$ is small or intermediate due to the flame stretch effects [52]. This is true for both homogeneous and heterogeneous flames. In particular, when $Le = 0.8$, $U$ monotonically decreases with $R_f$ for homogeneous and gaseous flames with $\delta = 0.1$, because of the continuously reduced enhancement effects from the positive stretch. However, when $\delta$ is further increased to 0.2, monotonic increase of $U$ with $R_f$ is observed. This implies that the fuel vapour addition dominates the foregoing stretch effects [51]. For relatively large $R_f$ (e.g. $10^3$), the propagation speed for homogeneous flame is independent on $Le$, similar to gaseous flame. However, $U$ for heterogeneous flame in large $R_f$ is appreciably affected by $Le$, due to the additional transport phenomenon related to droplet evaporation behind the flame front. This is consistent with the recent finding of planar spray heterogeneous flames with fuel mists [25].

Figure 12(b) shows the Lewis number effects on flame propagation speed when the evaporation heat exchange coefficient $\Omega$ varies. For homogeneous and heterogeneous flames, their tendencies for $U$ variations with $Le$ are generally similar to the counterpart flames in Fig. 12(a). When $Le = 0.8$, the kinetic contributions from fuel sprays exceed the flame stretch effects, which makes $U$ monotonically increase with $R_f$. This is observable for $\Omega = 0.1$ and 0.4, and different from the gaseous flame results.

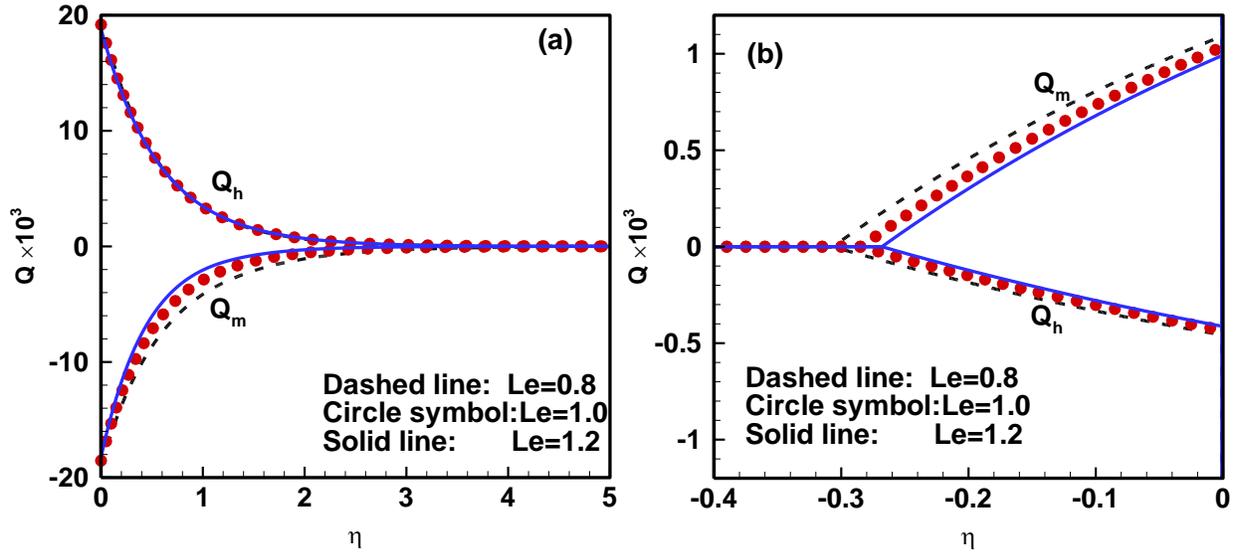

Fig. 13. Distributions of heat and mass fluxes for different Lewis numbers in (a) pre- and (b) post-flame zones. $q_v = 0.4$ and $R_f = 100$. The flame front lies at $\eta = 0$.

Figures 13(a) and 13(b) respectively show the heat and mass fluxes in pre- and post-flame zones of heterogeneous flames (marked with circles in Fig. 12) with $Le = 0.8$, 1.0 and 1.2. It is seen that in both zones variation of the Lewis number only affects the mass fluxes. Specifically, lower $Le$ increases the mass flux magnitude and hence enhances the flame reactivity, characterized by higher propagation speed shown in Fig. 12(a).

**4.5 Markstein length**

The effects of stretch rate on homogeneous and heterogeneous spherical spray flames will be further studied in this Section. For weakly stretched spherical flames ($R_f$>>1 or $K$<<1), the following relation holds between flame propagation speed $U$ and stretch rate $K$ [31,52]

$$U = U^0 - L \cdot K, \tag{54}$$

where $U^0$ is the flame speed at zero stretch rate. For spherical flames, stretch rate $K$ can be derived from $K = 2U/R_f$. Specifically, for gaseous and homogeneous spherical flames, the flame stretch rate $K$ is positive. Interestingly, for heterogeneous flames, due to the concurrent transport before and after the propagating flame front, $K$ is respectively positive and negative for them. Therefore, their individual effects, together with Lewis number, on flame propagation speeds would be opposite due to the reversed flame surface curvature, and therefore competitive. For instance, if $Le < 1$, the positive $K$ (for the unburned area) accelerates the flame, whereas the negative $K$ (for the burned area) decelerates it [52]. Therefore, Eq. (54) measures the gross response of the heterogeneous flame propagation speed to the stretch rate. Nonetheless, from Figs. 10, 11 and 13, it is seen that the transport in the burned area is comparatively lower, and therefore the mass and thermal diffusion in the fresh gas still dominate. Figure 14 shows the $U-K$ curves corresponding to the results in Fig. 12. For outwardly propagating spherical flames, the larger the flame radius, the lower the stretch rate. Figure 14 indicates that a linear relation exists between $U$ and $K$ when $K$ is low [24,53], and therefore the Markstein length $L$ (slope of the $U-K$ curve) can be calculated with Eq. (54).

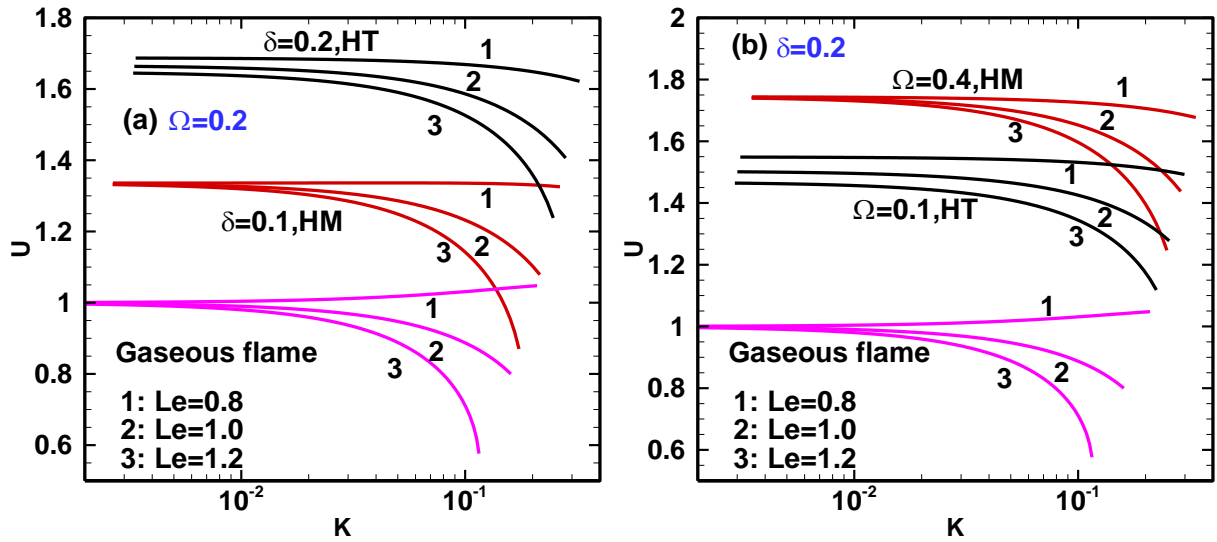

Fig. 14. Changes of flame propagation speed with stretch rate for different Lewis numbers when (a) $\Omega = 0.2$ and (b) $\delta = 0.2$.

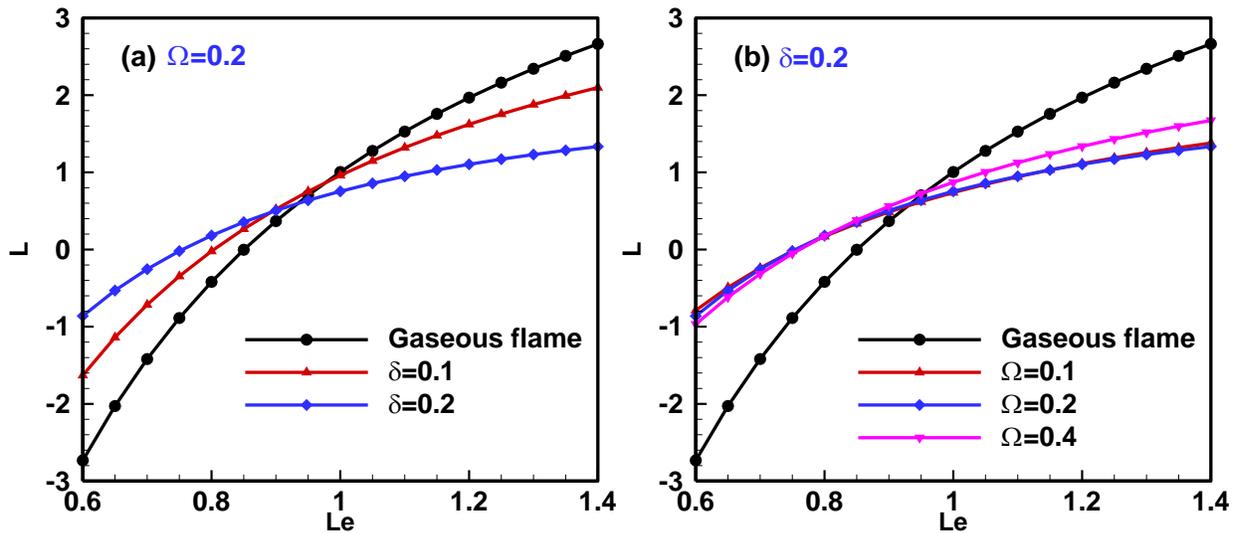

Fig. 15. Markstein length versus Lewis number when (a) $\Omega = 0.2$ and (b) $\delta = 0.2$.

Figure 15 shows the Markstein length with various Lewis numbers in the flames of Figs. 12 and 14. Generally, the Markstein length $L$ is considerably affected by droplet mass loading and evaporation heat exchange coefficient when the Lewis number deviates from unity. For $Le < 1$ ($Le > 1$), $L$ is negative (positive), indicating the flame propagation speed is enhanced (decreased) relative to $U^0$. The norms of $L$ in spray flames are reduced compared to gaseous flames with the same $Le$, which implies that the degrees of the enhancement or reduction of spray flame propagation are weakened. In addition, this extent is enhanced with the droplet loading $\delta$ when $\Omega$ is fixed, as shown in Fig. 15(a). However, it is shown from Fig. 15(b) that for fixed $\delta$, the sensitivity of the Markstein length to $\Omega$ is less pronounced.

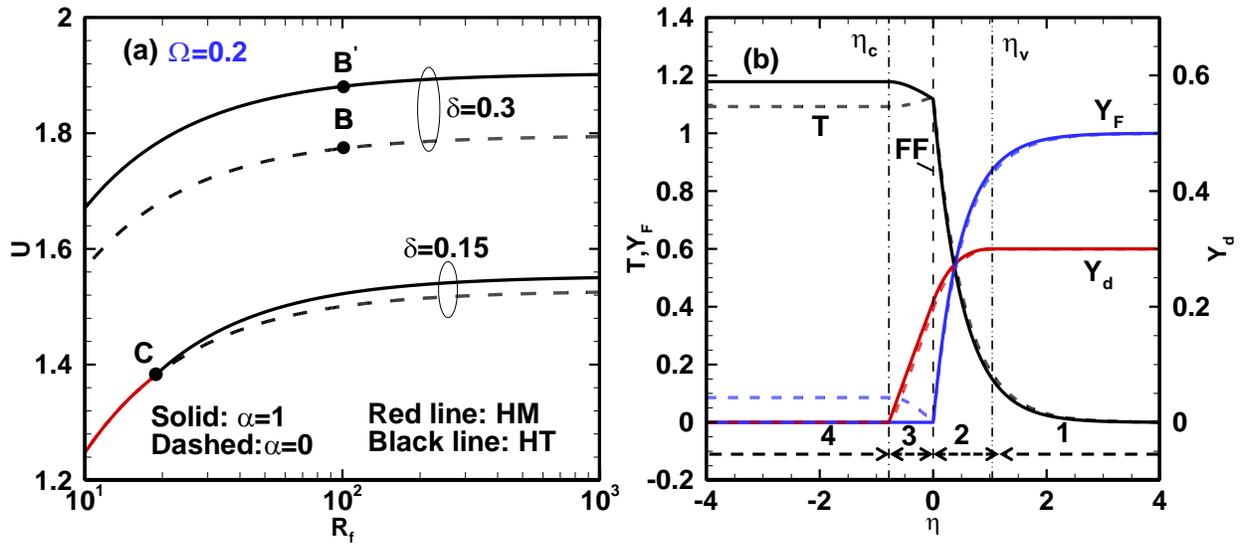

Fig. 16. (a) Changes of heterogeneous flame propagation speed with flame radius, (b) spatial distributions of temperature, fuel mass fraction and droplet mass loading for points B' (solid) and B (dashed). $Le = 1$.

### 4.6 Droplet burning behind a heterogeneous flame

In Sections 4.1−4.5, for heterogeneous flames, droplet burning behind the flame front is not considered. To examine this effect, two heterogeneous flames ($\delta = 0.15$ and 0.3) from Fig. 2(a) are re-calculated with $\alpha = 1$ in this Section. The corresponding $U - R_f$ curves are shown in Fig.

16(a). Apparently, for δ = 0.3, the heterogeneous flame propagation speed is higher when droplet burning is included. This is also reported by Han and Chen [24]. The structure of heterogeneous flame with droplet burning (B' in Fig. 16a, $R_f$ = 100) are shown in Fig. 3(b). For comparison, the results without droplet burning (B in Fig. 16a, also shown in Fig. 3b) are also added. Firstly, due to post-flame droplet burning, $Y_F$ in the post-flame zone is zero, since $\widetilde{\bar{\omega}} = \alpha \widetilde{\omega}_v$ in Eq. (2). Moreover, the heat release of droplet burning in the post-flame zone leads to higher temperature compared to the same condition when post-flame burning is not considered. Meanwhile, different from flame B, a negative temperature gradient exists behind the reaction front in flame B', resulting in higher propagation speed unveiled in Fig. 16(a).

Moreover, in Fig. 16(a), for δ = 0.15, a transition from homogeneous to heterogeneous flame occurs at C for both $\alpha$ = 0 and 1. Note that their transition point is the same. This is reasonable since droplet burning does not affect the results of homogeneous one. Nonetheless, when $\alpha$ = 1, sudden flame acceleration is observable from increased $dU/dR_f$ across C (not shown here), leading to slightly higher $U$ of heterogeneous flames with droplet burning.

Figure 17 shows the influences of droplet burning on Markstein length of heterogeneous flames with different Lewis numbers and droplet mass loadings. It is found that when δ = 0.3, droplet burning in the post-flame zone negligibly affect the Markstein length. However, when δ is reduced to 0.15, the magnitude of Markstein length is higher when $\alpha$ = 1, indicating that flame stretch has stronger effects on the propagation speeds when droplet burning is included. For heterogeneous flames with smaller droplet loading, the evaporation completion location is closer to the flame front (see Fig. 6a) and hence the burning droplets would more appreciably affect the leading flame front through modulating the local heat and mass transfer.

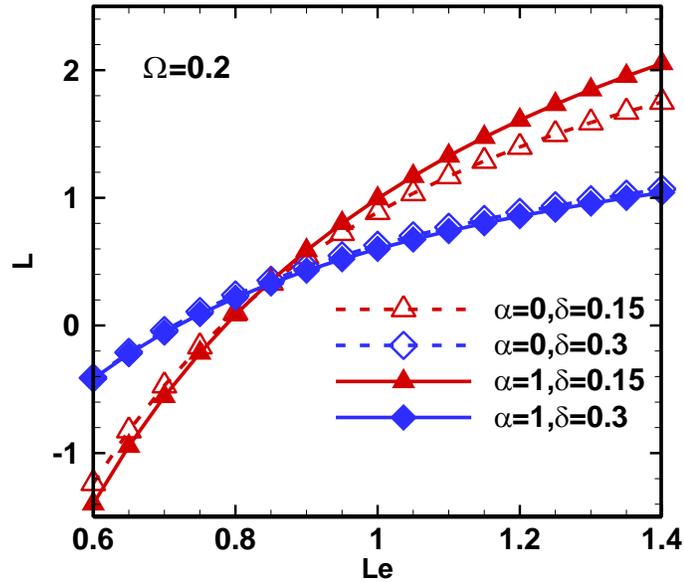

Fig. 17. Effects of droplet burning on Markstein length with various Lewis numbers and droplet mass loadings.

## 5 Conclusion

Propagation of weakly stretched premixed spray flames in localized homogeneous and heterogeneous reactants is investigated in this work. A general theory based on one-dimensional spherical spray flames with evaporating fuel droplets is developed and correlations are derived to describe flame propagation speed, flame temperature, droplet evaporation onset and completion locations. The theory enables the analysis on the influences of liquid fuel and gaseous mixture properties on propagation of spherical spray flames. The main conclusions are listed as below:

1. The flame propagation speed is enhanced with increased droplet mass loading and/or evaporation heat exchange coefficient (or evaporation rate). Opposite tendencies are observed when the latent heat is high, due to strong evaporation heat loss. For heterogeneous flame, there exist gradients for fuel vapor and gaseous temperature in the post-flame zone when post-flame burning is not considered. For large latent heat, when the droplet loading is large, flame bifurcation phenomenon occurs.

2. The evaporation completion location relies on the droplet properties. For large droplet loading or small heat exchange coefficient, the reactants around the flame front are heterogeneous. In some extreme condition, the droplets are distributed in the entire post-flame zone. Besides, the spray flame experiences a transition from localized homogeneous to heterogeneous conditions in intermediate droplet loading or heat exchange coefficient. These dynamic droplet behaviors can be predicted with the developed theory.

3. Heat and mass diffusion for both heterogeneous and homogeneous spray flame are considerably affected by presence of fuel droplets. In the pre-flame zone, they are enhanced (suppressed) with small (large) latent heat of the liquid fuel. Besides, there exist heat and mass diffusion in the post-flame evaporation zone when post-flame burning is not considered.

4. The spray flame propagation speed decreases with Lewis number. For heterogeneous flame, the flame speeds at large flame radius is considerably affected by Lewis number due to the continuous heat and mass diffusion in the post-flame zone. When Lewis number is less than unity, the monotonicity of flame propagation speed variation is influenced by droplet loading and/or evaporation rate. Besides, the mass flux magnitude increases with Lewis number in both pre- and post-flame zones.

5. The Markstein length is affected by droplet loading and evaporation heat exchange coefficient when the Lewis number deviates from unity. Their magnitudes are reduced compared to gaseous flames, which indicates that the degrees of the enhancement or reduction of flame propagation by stretch rate are weakened. Conversely, sensitivity of the Markstein length to the heat exchange coefficient is less pronounced.

6. For heterogeneous flames with post-flame droplet burning, the heat release due to droplet burning leads to higher burned gas temperature and flame propagation speed. The Markstein length is more pronouncedly affected by droplet burning when the liquid fuel loading is relatively low.

**Acknowledgement**

QL is supported by NUS Research Scholarship. HZ thanks Professor Zheng Chen at Peking University for helpful discussion.